\documentclass[aps,prb,showpacs,twocolumn,superscriptaddress]{revtex4-1}%
\usepackage{amssymb}
\usepackage{epsfig}
\usepackage{epsf}
\usepackage[usenames]{color}
\usepackage{upgreek}
\usepackage{amsmath}
\usepackage{xcolor}
\usepackage{hyperref}
\usepackage{amsfonts}
\usepackage{graphicx}
\begin{document}

\title{Nonlocal magnon-polaron transport in yttrium iron garnet}
\author{L.J. Cornelissen}
\thanks{These authors contributed equally to this work}
\affiliation{Physics of Nanodevices, Zernike Institute for Advanced Materials, University
	of Groningen, Nijenborgh 4, 9747 AG Groningen, The Netherlands }

\email{l.j.cornelissen@rug.nl}
\author{K. Oyanagi}
\thanks{These authors contributed equally to this work}
\affiliation{Institute for Materials Research, Tohoku University, Sendai 980-8577, Japan}

\author{T. Kikkawa}
\affiliation{Institute for Materials Research, Tohoku University, Sendai 980-8577, Japan}
\affiliation{WPI Advanced Institute for Materials Research, Tohoku University, Sendai 980-8577, Japan}

\author{Z. Qiu}
\affiliation{WPI Advanced Institute for Materials Research, Tohoku University, Sendai 980-8577, Japan}
\author{T. Kuschel}
\affiliation{Physics of Nanodevices, Zernike Institute for Advanced Materials, University
	of Groningen, Nijenborgh 4, 9747 AG Groningen, The Netherlands }
\author{G.E.W. Bauer}
\affiliation{Physics of Nanodevices, Zernike Institute for Advanced Materials, University
	of Groningen, Nijenborgh 4, 9747 AG Groningen, The Netherlands }
\affiliation{Institute for Materials Research, Tohoku University, Sendai 980-8577, Japan}
\affiliation{WPI Advanced Institute for Materials Research, Tohoku University, Sendai 980-8577, Japan}
\affiliation{Center for Spintronics Research Network, Tohoku University, Sendai 980-8577, Japan}
\author{B.J. van Wees}
\affiliation{Physics of Nanodevices, Zernike Institute for Advanced Materials, University
	of Groningen, Nijenborgh 4, 9747 AG Groningen, The Netherlands }
\author{E. Saitoh}
\affiliation{Institute for Materials Research, Tohoku University, Sendai 980-8577, Japan}
\affiliation{WPI Advanced Institute for Materials Research, Tohoku University, Sendai 980-8577, Japan}
\affiliation{Center for Spintronics Research Network, Tohoku University, Sendai 980-8577, Japan}
\affiliation{Advanced Science Research Center, Japan Atomic Energy Agency, Tokai 319-1195, Japan}

\begin{abstract}
The spin Seebeck effect (SSE) is observed in magnetic insulator\textbar{}heavy metal bilayers as an inverse spin Hall effect voltage under a temperature gradient. The SSE can be detected nonlocally as well, viz. in terms of the voltage in a second metallic contact (detector) on the magnetic film, spatially separated from the first contact that is used to apply the temperature bias (injector). Magnon-polarons are hybridized lattice and spin waves in magnetic materials, generated by the magnetoelastic interaction. Kikkawa \emph{et al.} [Phys. Rev. Lett. \textbf{117}, 207203 (2016)] interpreted a resonant enhancement of the local SSE in yttrium iron garnet (YIG) as a function of the magnetic field in terms of magnon-polaron formation. Here we report the observation of magnon-polarons in \emph{nonlocal} magnon spin injection/detection devices for various injector-detector spacings and sample temperatures. Unexpectedly, we find that the magnon-polaron resonances can suppress rather than enhance the nonlocal SSE. Using finite element modelling we explain our observations as a competition between the SSE and spin diffusion in YIG. These results give unprecedented insights into the magnon-phonon interaction in a key magnetic material.
%Magnon-polarons are composite quasiparticles, arising due to strong magnetoelastic coupling in magnetic systems. This coupling causes magnons and phonons to hybridize, forming anticrossings in their dispersion relations for appropriate magnitudes of the external magnetic field. We investigated magnon-polaron transport in yttrium iron garnet using nonlocal magnon spin injection/detection devices. The signature of magnon-polarons observed so far is the resonant enhancement of the local spin Seebeck signal as the magnetic field crosses the threshold value for magnon-phonon hybridization. Measured nonlocally however, the magnon-polaron resonance can take the form of either resonant signal \emph{enhancement} or signal \emph{suppression}, depending on injector-detector spacing and sample temperature. Using a finite element model of our devices, we can qualitatively explain our observations by noting that the hybridization not only affects the bulk spin Seebeck coefficient, which relates the magnon spin current to the applied temperature gradient, but also the magnon spin conductivity. Whether suppression or enhancement occurs depends on the ratio in which the magnon spin conductivity and spin Seebeck coefficient are affected by the hybridization.
\end{abstract}

\maketitle
When sound travels through a magnet the local distortions of the lattice exert torques on the magnetic order due to the magnetoelastic coupling\cite{Kittel1958}. By reciprocity, spin waves in a magnet affect the lattice dynamics. The coupling between spin and lattice waves (magnons and phonons) has been intensively researched in the last half century\cite{Eshbach1963, Schlomann1964}. Yttrium iron garnet (YIG) has been a singularly useful material here, because it can be grown with exceptional magnetic and acoustic quality\cite{Eshbach1963}. Magnons and phonons hybridize at the (anti)crossing of their dispersion relations, a regime that has attracted recent attention\cite{PhysRevB.89.184413, Ogawa2015, Kamra2015,Shen2015, Guerreiro2015, Kikkawa2016, Flebus2017}. When the quasiparticle lifetime-broadening is smaller than the interaction strength, the strong coupling regime is reached; the resulting fully mixed quasiparticles have been referred to as magnon-polarons\cite{Kamra2015,Shen2015}.

In spite of the long history and ubiquity of the magnon-phonon interaction, it still leads to surprises. Evidence of a sizeable magnetoelastic coupling in YIG was recently found in experiments on spin caloritronic effects, i.e. the spin Peltier\cite{Flipse2014} and spin Seebeck effect\cite{Agrawal2013,PhysRevB.88.094410} (SPE and SSE respectively). Recently, Kikkawa \emph{et al.} showed that the hybridization of magnons and phonons can lead to a resonant enhancement of the local SSE in YIG\cite{Kikkawa2016}. Bozhko \emph{et al.} found that this hybridization can play a role in the thermalization of parametrically excited magnons using Brillouin light scattering. They observed an accumulation of magnon-polarons in the spectral region near the anticrossing between the magnon and transverse acoustic phonon modes\cite{Bozhko2016}. However, these previous experiments did not address the transport properties of magnon-polarons.

Nonlocal spin injection and detection experiments are of great importance in probing the transport of spin in metals\cite{Jedema2001}, semiconductors\cite{Lou2006} and graphene\cite{Tombros2007}. Varying the distance between the spin injection and detection contacts allows for the accurate determination of the transport properties of the spin information carriers in the channel, such as the spin relaxation length\cite{Fabian2004}. Recently, it was shown that this kind of experiments are not limited to (semi)conducting materials, but can also be performed on magnetic insulators\cite{Cornelissen2015}, where the spin information is carried by magnons. Such nonlocal magnon spin transport experiments have provided additional insights in the properties of magnons in YIG, for instance by studying the transport as a function of temperature\cite{Goennenwein2015, Ganzhorn2017, Zhou2017, Cornelissen2016b} or external magnetic field\cite{Cornelissen2016}. Finally, the nonlocal magnon spin injection/detection scheme can play a role in the development of efficient magnon spintronic devices, for example magnon based logic gates\cite{Chumak2015,Ganzhorn2016a}. In this study, we make use of nonlocal magnon spin injection and detection devices to investigate the transport of magnon-polarons in YIG. 

Magnons can be excited magnetically using the oscillating magnetic field generated by a microwave frequency ac current\cite{Chumak2015}, or electrically using a dc current in an adjacent material with a large spin Hall angle, such as platinum\cite{Cornelissen2015}. Finally, they can be generated thermally by the SSE\cite{Uchida2010, Vlietstra2014, Weiler2012, Meier2013}, in which a thermal gradient in the magnetic insulator drives a magnon spin current parallel to the induced heat current. 

The generation of magnons via the SSE can be detected in several configurations: First, the heater-induced configuration (hiSSE)\cite{Uchida2010a}, which consists of a bilayer YIG\textbar{}heavy metal sample that is subject to external Peltier elements to apply a temperature gradient normal to the plane of the sample. The SSE then generates a voltage across the heavy metal film (explained in more detail below), which can be recorded. Second, the current-induced configuration (ciSSE)\cite{Schreier2013,Vlietstra2014} in which the heavy metal detector used to detect the SSE voltage is simultaneously used as a heater. A current is sent through the heavy metal film, creating a temperature gradient in the YIG due to Joule heating. Due to this temperature gradient, the SSE generates a voltage across the heavy metal film, which can again be recorded. Third, the nonlocal SSE (nlSSE)\cite{Cornelissen2015, Shan}, in which a current is sent through a narrow heavy metal strip to generate a thermal gradient via Joule heating as well. However, the SSE signal resulting from this thermal gradient is detected in a second heavy metal strip, located some distance away from the injector. 

In the nlSSE, the magnons responsible for generating a signal in the detector strip are generated in the injector vicinity and then diffuse through the magnetic insulator to the detector. The temperature gradient underneath a detector located several microns to tens of microns from the injector does not contribute significantly to the measured voltage\cite{Cornelissen2016a, Cornelissen2016b}. In contrast, the hiSSE and ciSSE always have a significant temperature gradient directly underneath the detector. The hiSSE and ciSSE are therefore local SSE configurations, contrary to the nlSSE which is nonlocal.

In all three configurations, the resulting voltage across the heavy metal film is due to magnons which are absorbed at the YIG\textbar{}detector interface, causing spin-flip scattering of conduction electrons and generating a spin current and spin accumulation in the detector. Due to the inverse spin Hall effect\cite{Saitoh2006}, this spin accumulation is converted into a charge voltage that is measured. 

At specific values for the external magnetic field, the phonon dispersion is tangent to that of the magnons and the magnon and phonon modes are strongly coupled over a relatively large region in momentum space. At these resonant magnetic field values, the effect of the magnetoelastic coupling is at its strongest and magnon-polarons are formed efficiently. If the acoustic quality of the YIG film is better than the magnetic one, magnon-polaron formation leads to an enhancement in the hiSSE signal at the resonant magnetic field\cite{Kikkawa2016}. This enhancement is attributed to an increase in the effective bulk spin Seebeck coefficient $\zeta$, which governs the generation of magnon spin current by a temperature gradient in the magnet. This was demonstrated experimentally by measuring the spin Seebeck voltage in the hiSSE configuration\cite{Kikkawa2016}, establishing the role of magnon-polarons in the thermal generation of magnon spin current.

Here we make use of the nlSSE configuration to directly probe not only the generation, but also the transport of magnon-polarons. We show that in the YIG samples under investigation not only $\zeta$, but also the magnon spin conductivity $\sigma_m$ is resonantly enhanced by the hybridization of magnons and phonons, which leads to signatures in the nonlocal magnon spin transport signals clearly distinct from the hiSSE observations. Notably, resonant features in nonlocal transport experiments have very recently been theoretically predicted by Flebus \emph{et al.}\cite{Flebus2017}, who calculated the influence of magnon-polarons on the YIG transport parameters such as the magnon spin and heat conductivity and the magnon spin diffusion length. 

\section{Results}
\label{sec:results}
\subsection{Sample characteristics}
\label{subsec:samplecharacterstics}
Our nonlocal devices consist of multiple narrow, thin platinum strips (typical dimensions are $100$ $\mu$m $\times$ $100$ nm $\times$ $10$ nm [$\mathrm{l\times w \times t}$]) deposited on top of a YIG thin film and separated from each other by a centre-to-centre distance $d$. We have performed measurements of nonlocal devices on YIG films from Groningen and Sendai, both of which are grown by liquid phase epitaxy on a GGG substrate. The YIG film thickness is 210 nm (2.5 $\mathrm{\mu m}$) for YIG from Groningen (Sendai). In Sendai, five batches of devices where investigated (sample S1 to S4) on pieces cut from the same YIG wafer. In Groningen, two batches of devices where investigated (G1 and G2). The platinum strips are contacted using Ti/Au contacts (see Methods for fabrication details) Figure \ref{fig:figure1}a shows an optical microscope image of a typical device, with the electrical connections indicated schematically. The central strip functions as a magnon injector while the two outer strips are magnon detectors, measuring the nonlocal signal at different distances from the injector. 

\subsection{Experimental results}
\label{subsec:experimentalresults}
A nonlocal signal is generated in the devices by passing a low frequency ac current $I(\omega)$ (typically $w/(2\pi)<20$ Hz and $I_{\mathrm{rms}}=100$ $\mathrm{\mu}$A) through the injector. This leads to both thermal and electrical generation of magnons, as outlined above. The voltage that is due to the thermally generated magnons is proportional to the excitation current squared, and hence can be directly detected in the second harmonic detector response $V(2\omega)$ (i.e. the voltage measured at twice the excitation frequency). Simultaneously, the voltage due to electrically generated magnons can be measured in the first harmonic response $V(1\omega)$\cite{Cornelissen2015}. The sample is placed in an external magnetic field $H$, under an angle $\alpha=90^{\circ}$ to the injector/detector strips. 

Figure \ref{fig:figure1}b shows the results of two typical nonlocal measurements at different distances, in which $\mu_0 H$ is varied from $-3.0$ to $3.0$ T. Several distinct features can be seen in these results. As the magnetic field is swept through zero, the YIG magnetization and hence the magnon spin polarization change direction, since a magnon always carries a spin opposite to the majority spin in the magnet. This causes a reversal of the polarization of the spin current absorbed by the detector and consequently the voltage $V_{\mathrm{nlSSE}}$ changes sign. Additionally, $V_{\mathrm{nlSSE}}$ for short distance $d$ (Figure~\ref{fig:figure1}b bottom panels) shows an opposite sign compared to $V_{\mathrm{nlSSE}}$ for long distance (Figure~\ref{fig:figure1}b top panels). This sign-reversal for short distances is a characteristic feature of the nlSSE\cite{Cornelissen2015} that has so far been observed to depend on both the thickness of the YIG film $t_{\mathrm{YIG}}$ (roughly speaking, at room temperature when $d<t_{\mathrm{YIG}}$ the sign will be opposite to that for $d>t_{\mathrm{YIG}}$\cite{Shan}) as well as the sample temperature, where a lower temperature reduces the distance at which the sign-change occurs\cite{Ganzhorn2017,Zhou2017}. 

The sign for short distances corresponds to the sign one obtains when measuring the SSE in its local configurations (hiSSE, indicated schematically in Figure~\ref{fig:figure1}c or ciSSE). The results for a hiSSE measurement on sample S3 as a function of $H$ are shown in Figure~\ref{fig:figure1}d, and $V_{\mathrm{hiSSE}}$ clearly shows the same sign as $V_{\mathrm{nlSSE}}$ for short distance. We will discuss the origin of this sign-change in more detail later in this manuscript. The data shown in Figure~\ref{fig:figure1} are from samples with $t_{\mathrm{YIG}}=2.5$ $\mathrm{\mu}$m, hence the different signs for $d=2$ $\mathrm{\mu}$m and $d=6$ $\mathrm{\mu}$m.

Resonant features can be observed in the data for $|\mu_0 H|=\mu_0 H_{\mathrm{TA}}\approx2.3$ T, where the subscript TA signifies that these features stems from the hybridization of magnons with phonons in the transverse acoustic mode, rather than the longitudinal acoustic mode (LA) which is expected at larger magnetic fields. The rightmost panels of Figure~\ref{fig:figure1} show a close-up of the data around $H=H_{\mathrm{TA}}$. For small $d$ the magnon-phonon hybridization causes a resonant \emph{enhancement} (the absolute value is increased) of $V_{\mathrm{nlSSE}}$, while for large $d$ a resonant \emph{suppression} (the absolute value is reduced) occurs. 

Figure~\ref{fig:first_second_harmonic} shows the results of a magnetic field sweep from sample G1 for both electrically generated magnons (first harmonic) and thermally generated magnons (second harmonic). A feature at $|H|=H_{\mathrm{TA}}$ can be resolved both in the first and second harmonic voltage. This suggests that magnon-phonon hybridization does not only affect the YIG spin Seebeck coefficient, as the first harmonic signal is generated independent of $\zeta$. It indicates that not only the generation, but also the transport of magnons is affected by the hybridization. In the second harmonic, the signal is clearly suppressed at the resonant magnetic field. Unfortunately, because the feature in the first harmonic is barely larger than the noise floor in the measurements (see Fig.~\ref{fig:first_second_harmonic}a and inset), we cannot conclude whether the signal due to electrical magnon generation is enhanced or suppressed at the resonance. Due to the fact that the effect in the first harmonic is so small, in the remainder of this paper we present a systematic study of the effect in the second harmonic, the nlSSE.

The resonant magnetic fields are different for the TA and LA modes ($H_{\mathrm{TA}}$ and $H_{\mathrm{LA}}$, respectively). Due to the higher sound velocity in the LA phonon mode, $H_{\mathrm{TA}}<H_{\mathrm{LA}}$, and the resonance due to magnons hybridizing with phonons in this mode can also be observed in our nonlocal experiments. In the Supplementary Material (section A) we show the results of a magnetic field scan over an extended field range, and it can be seen that the resonance at $H_{\mathrm{LA}}$ also causes a suppression of the nlSSE signal, similar to the $H_{\mathrm{TA}}$ resonance. This is comparable to the case for the hiSSE configuration, in which the $H_{\mathrm{LA}}$ and $H_{\mathrm{TA}}$ resonances both show similar behaviour in the sense that they both \emph{enhance} the hiSSE signal. For the nlSSE case at distances larger than the sign-change distance, both resonances \emph{suppress} the signal.

We now focus on the resonance at $H_{\mathrm{TA}}$ in the nlSSE data and carried out nonlocal measurements as a function of magnetic field for various temperatures and distances. Figure~\ref{fig:figure2}a (b) shows the distance (temperature) dependent results, obtained from sample S1 (S2). The regions where the sign of the nlSSE equals that of the hiSSE are shaded blue. From Figure~\ref{fig:figure2}a the sign-change in $V_{\mathrm{nlSSE}}$ can be clearly seen to occur between $d=2$ and $d=5$ $\mu$m, as at $d=2$ $\mu$m the nlSSE sign is equal to that of the hiSSE for any value of the magnetic field, whereas for $d=5$ $\mu$m it is opposite. Additionally, when comparing the $V_{\mathrm{nlSSE}}-H$ curves for 300 K and 100 K in Figure~\ref{fig:figure2}b, the effect of the sample temperature on the sign-change is apparent: At 100 K, the nlSSE sign is opposite to that of the hiSSE over the whole curve. Furthermore, Figure~\ref{fig:figure2}b demonstrates the influence of the magnetic field on the sign change, for instance in the curve for $T=160$ K. At low magnetic fields, the nlSSE sign still agrees with the hiSSE sign (inside the blue shaded region), but around $|\mu_0H|=1.5$ T the signal changes sign.

In addition, Figure~\ref{fig:figure2}a shows that the role of the magnon-polaron resonance changes as the nlSSE signal undergoes a sign change. For $d\leq2$ $\mu$m, magnon-phonon hybridization enhances $V_{\mathrm{nlSSE}}$ at $H=H_{\mathrm{TA}}$, whereas for $d\geq5$ $\mu$m $V_{\mathrm{nlSSE}}$ is suppressed at the resonance magnetic field. Similarly, from Figure~\ref{fig:figure2}b we observe that at temperatures $T>160$ K, magnon-phonon hybridization enhances the nlSSE signal at $H=H_{\mathrm{TA}}$, while at $T\leq160$ K the nlSSE is suppressed at $H_{\mathrm{TA}}$. Since the thermally generated magnon spin current is related to the thermal gradient by $\mathbf{j}_m\propto-\zeta\nabla T$, a resonant enhancement in $\zeta$ should lead to an enhancement of the nlSSE signal at all distances and temperatures, which is inconsistent with our observations. This is a further indication that not only the generation, but also the transport of magnons is influenced by magnon-polarons. %In Sections~\ref{subsec:modellingresults} and \ref{subsec:comparison} we will further discuss the distance dependence of the nlSSE signal, in particular the mechanism behind the sign-change, and show that this can be qualitatively explained using numerical simulations of our devices.

The temperature dependence of the low-field amplitude of the nlSSE $V_{\mathrm{nlSSE}}^0$ and the magnitude of the resonance $V_{\mathrm{TA}}$ (defined in Figure~\ref{fig:figure1}b) are shown in Figure~\ref{fig:figure3}a and \ref{fig:figure3}b respectively. The curve for $V_{\mathrm{nlSSE}}^0$ at $d=6$ $\mathrm{\mu}$m agrees well with an earlier reported temperature dependence of the nlSSE at distances which are larger than the film thickness\cite{Cornelissen2016b}, while that at $d=2$ $\mathrm{\mu}$m qualitatively agrees with earlier reports for distances shorter than the YIG film thickness\cite{Ganzhorn2017,Zhou2017}. Moreover, from the distance dependence of $V_{\mathrm{nlSSE}}^0$ we have extracted the magnon spin diffusion length $\lambda_m$ as a function of temperature, which is shown in the Supplementary Material (section B). $\lambda_m(T)$ obtained from the Sendai YIG approximately agrees with that for Groningen YIG\cite{Cornelissen2016b} for temperatures $T>30$ K, but differs in the low temperature regime. For further discussion we refer to the Supplementary Material of this manuscript. The temperature dependence of $V_{\mathrm{TA}}$ is different from that of $V_{\mathrm{nlSSE}}^0$, since first of all no change in sign occurs here even for $d=2$ $\mu$m and furthermore a clear minimum appears in the curve around $T=50$ K. This indicates that the resonance has a different origin than the nlSSE signal itself, i.e. magnon-polarons are affected differently by temperature than pure magnons. 

The resonant magnetic field $H_{\mathrm{TA}}$ decreases with increasing temperature, reducing from $\mu_0H_{\mathrm{TA}}\approx2.5$ T at 3 K to $\mu_0H_{\mathrm{TA}}\approx2.2$ T at room temperature as shown in Figure~\ref{fig:figure3}c. In earlier work by some of us regarding the magnetic field dependence of the nonlocal magnon transport signal at room temperature, structure in the data at $\mu_0 H=2.2$ T was indeed observed\cite{Cornelissen2016}, but not understood at that time. It is now clear that this structure can be attributed to magnon-phonon hybridization. $H_{\mathrm{TA}}$ depends on the following three parameters\cite{Kikkawa2016}: The YIG saturation magnetization $M_s$, the spin wave stiffness constant $D_{\mathrm{ex}}$ and the TA-phonon sound velocity $c_{\mathrm{TA}}$. $D_{\mathrm{ex}}$ is approximately constant for $T<300$ K\cite{Lecraw1961} and both $M_s$ and $c_{\mathrm{TA}}$ decrease with temperature. The reduction of $H_{\mathrm{TA}}$ as temperature increases from 3 K to 300 K can be explained by accounting for a $7$ \% decrease of $c_{\mathrm{TA}}$ in the same temperature interval, taking the temperature dependence of $M_s$ into consideration\cite{Solt1962}. The results regarding the behaviour of the magnon-polaron resonance qualitatively agree for the Sendai and Groningen YIG (see Supplementary Material (section C) for the temperature dependent results for sample G2).

Moreover, we performed measurements of the nlSSE signal as a function of the injector current, and found that the nlSSE scales linearly with the square of the current at high temperatures, as expected. However, at low temperatures ($T<10$ K) and sufficiently high currents (typically, $I>50$ $\mathrm{\mu}$A), this linear scaling breaks down (see Supplementary Material (section D)). This could be a consequence of the strong temperature dependence of the YIG and GGG heat conductivity at these temperatures\cite{Daudin1982, PhysRevB.90.064421}. The injector heating causes a small increase in the average sample temperature which increases the heat conductivities of the YIG and GGG, thereby driving the system out of the linear regime. However, it might also be related to the bottleneck effect which is observed in parametrically excited YIG\cite{Bozhko2016}. A more detailed investigation is needed in order to establish the origin of the nonlinearity.

Finally, we have investigated the ciSSE configuration, meaning that current heating of the Pt injector is used to drive the SSE and the (local) voltage across the injector is measured. The sign of the ciSSE voltage corresponds to that obtained in the hiSSE configuration. However, no resonant features were observed in the ciSSE measurements, contrary to the hiSSE and nlSSE configurations. We believe that this is due to the low signal-to-noise ratio in the ciSSE configuration, which could cause the feature to be smaller than the noise level in our ciSSE measurements. We refer to the Supplementary Material (section E) for further discussion.

\subsection{Modelling}
\label{subsec:modellingresults}
The physical picture underlying the thermal generation of magnons has been a subject of debate in the magnon spintronics field recently. Previous theories explain the SSE as being due to thermal spin pumping, caused by a temperature difference between magnons in the YIG and electrons in the platinum\cite{PhysRevB.88.094410,Xiao2010,Adachi2013}. However, the recent observations of nonlocal magnon spin transport and the nlSSE give evidence that not only the interface but also the bulk magnet actively contributes and even dominates the spin current generation. At elevated temperatures the energy relaxation should be much more efficient than the spin relaxation, which implies that the magnon chemical potential (and its gradient) is more important as a non-equilibrium parameter than the temperature difference between magnons and phonons. A model for thermal generation of magnon spin currents based on the bulk SSE\cite{PhysRevB.89.014416} which takes into account a non-zero magnon chemical potential has been proposed in order to explain the observations\cite{Cornelissen2016a}. 

This model has been reasonably successful in explaining the nonlocal signals (due to both thermal and electrical generation) in the long distance limit\cite{Shan, Cornelissen2016b}, yet is not fully consistent with experiments in the short distance limit for thermally generated magnons\cite{Shan}. The model is explained in detail in Refs. \onlinecite{Cornelissen2016a, Shan}, and is described concisely in the Methods section of this manuscript. The physical picture captured by the model is explained in Figure~\ref{fig:figure5}a and b, where for this study we focus on the thermally generated magnons driving the nlSSE. In Figure~\ref{fig:figure5}a a schematic side-view of the YIG\textbar{}GGG sample with a platinum injector strip on top is shown. A current is passed through the injector, causing it to heat up to temperature $T_H$. The bottom of the GGG substrate is thermally anchored at $T_0$. As a consequence of Joule heating, a thermal gradient arises in the YIG, driving a magnon current $J^m_Q=-\zeta/T\nabla T$ parallel to the heat current, i.e. radially away from the injector. This reduces the number of magnons in the region directly below the injector (magnon depletion). 

In Figure~\ref{fig:figure5}b the same schematic cross-section is shown, but now the colour coding refers to the magnon chemical potential $\mu_m$. Directly below the injector contact $\mu_m$ is negative due to the magnon depletion in this region ($\mu^-$). At the YIG\textbar{}GGG interface, magnons accumulate since they are driven towards this interface by the SSE but are reflected by the GGG, causing a positive magnon chemical potential $\mu^+$ to build up. Note that the $\mu^-$ and $\mu^+$ regions are not equal in size since part of the magnon depletion is replenished by the injector contact, which acts as a spin sink. Due to the gradient in magnon chemical potential, a diffuse magnon spin current $J^m_d$ now arises in the YIG given by $J^m_d=-\sigma_m\nabla\mu_m$. 

The combination of these two processes leads to a typical magnon chemical potential profile as shown in Figure~\ref{fig:figure5}c, which is obtained from the finite element model (FEM) at room temperature. The sign change from $\mu^-$ to $\mu^+$ occurs at a distance of roughly $d_{sc}=2.6$ $\mu$m from the injector, comparable to the YIG film thickness. 

Here we used the effective spin conductance of the Pt\textbar{}YIG interface $g_s$ as a free parameter in order to get approximate agreement between the modelled and experimentally observed sign-change distance $d_{sc}$ (see Methods for the further details of the model). The value for $g_s$ is approximately a factor 30 lower than what we calculated from theory\cite{Cornelissen2016a} and used in our previous work\cite{Cornelissen2016b}. When using $g_s=9.6\times10^{12}$ S/m$^2$ as in previous work, $d_{sc}\approx300$ nm which is much shorter than what we observe in the experiments. This discrepancy between the models for electrically and thermally generated magnon transport might indicate that some of the material parameters such as spin or heat conductivity and spin diffusion length (for both YIG and platinum) we use are not fully accurate. However, it is also conceivable that the models are not complete and need to be refined further\cite{Shan}, for instance by including temperature difference at material interfaces which are currently neglected.

The value of $d_{sc}$ depends mainly on four parameters: The thickness of the YIG film $t_{\mathrm{YIG}}$, the transparency of the platinum\textbar{}YIG injector interface, parameterized in the effective spin conductance $g_s$, the magnon spin conductivity of the YIG $\sigma_m$ and finally the magnon spin diffusion length $\lambda_m$. At high temperatures (i.e. close to room temperature), the thermal conductivities $\kappa_{\mathrm{GGG}}$ and $\kappa_{\mathrm{YIG}}$ are similar in magnitude\cite{Slack1971} and affect $d_{sc}$ only weakly, allowing us to focus here on the spin transport.

Increasing $t_{\mathrm{YIG}}$ or $\sigma_m$ increases $d_{sc}$ since this reduces the spin resistance of the YIG film, allowing the depleted region to spread further throughout the YIG. However, increasing $g_s$ or $\lambda_m$ causes the opposite effect and reduces $d_{sc}$ since this increases the amount of $\mu^-$ which is absorbed by the injector contact compared to that which relaxes in the YIG. The precise dependency of $d_{sc}$ on these parameters is nontrivial but can be explored using our finite element model. Ganzhorn \emph{et al.} and Zhou \emph{et al.} in Refs.~\onlinecite{Ganzhorn2017,Zhou2017} observed that $d_{sc}$ becomes smaller with lower temperatures. This indicates that the ratio of the effective spin resistance of YIG to that of the Pt contact increases, causing spins to relax preferentially into the contact and thereby reducing the extend of $\mu^-$. 

Flebus \emph{et al.} developed a Boltzmann transport theory for magnon-polaron spin and heat transport in magnetic insulators\cite{Flebus2017}. Here we implement the salient features of magnon-polarons into our finite element model. We observe that when the combination of $g_s$, $\lambda_m$, $\sigma_m$, $t_{\mathrm{YIG}}$ and $d$ is such that the detector is probing the depletion region, i.e. $\mu^-$, the magnon-polaron resonance causes enhancement of the nlSSE signal. Conversely, when the detector is probing $\mu^+$ the resonance causes a suppression of the signal. This cannot be explained by assuming that the only effect of the magnon-polaron resonance is the enhancement of $\zeta$, as this would simply increase the thermally driven magnon spin current $J^m_Q$ and hence enhance both $\mu^-$ and $\mu^+$. To understand this behaviour, we have to account for the enhancement of $\sigma_m$ by the magnon-polaron resonance as well. 

A resonant increase in $\sigma_m$ leads to an increased diffusive backflow current $J^m_d$, which can lead to a reduction of the magnon spin current reaching the detector at large distances. We model the effect of the magnon-phonon hybridization by assuming a field-dependent magnon spin conductivity $\sigma_m(H)$ and bulk spin Seebeck coefficient $\zeta(H)$, which are both enhanced at the resonant field $H_{\mathrm{TA}}$. Note that the field-dependence only includes the contribution from the magnon-polarons\cite{Flebus2017}, and does not include the effect of magnons being frozen out by the magnetic field\cite{Kikkawa2015, Cornelissen2016, Jin2015, Kikkawa2016a} since this is not the focus of this study. The parameter values used in the model are given in the Methods section of this paper. The model is used to calculate the spin current flowing into the detector contact as a function of magnetic field, from which we calculate the voltage drop over the detector due to the inverse spin Hall effect. We then vary the ratios of enhancement for $\sigma_m$ and $\zeta$, i.e. $f_{\sigma}=\sigma_m(H_\mathrm{TA})/\sigma_m^0$ and $f_{\zeta}=\zeta(H_{\mathrm{TA}})/ \zeta^0$, where $\sigma_m^0$ and $\zeta^0$ are the zero field magnon spin conductivity and spin Seebeck coefficient and $\sigma_m(H_\mathrm{TA}), \, \zeta(H_{\mathrm{TA}})$ are these parameters at the resonant field. The ratio of enhancement $\delta=f_{\zeta}/f_{\sigma}$ is crucial in obtaining agreement between the experimental and modelled data. To change delta, we fix $f_{\zeta}=1.09$ and vary $f_{\sigma}$. The value for $f_{\zeta}$ is comparable to the enhancement in $\zeta$ calculated from theory for low temperatures\cite{Flebus2017}.

\subsection{Comparison between model and experiment}
\label{subsec:comparison}
Figure~\ref{fig:distance_dependence} shows a comparison between the distance dependence of $V_{\mathrm{nlSSE}}^0$ and $V_{\mathrm{TA}}$ obtained from experiments (Fig.~\ref{fig:distance_dependence}a) and the finite element model (Fig.~\ref{fig:distance_dependence}b and c) at room temperature. In Figure~\ref{fig:distance_dependence}a, $V^0_{\mathrm{nlSSE}}$ shows a change in sign around $d=4\,\mu$m, while $V_{\mathrm{TA}}$ has a positive sign over the whole distance range. Fig.~\ref{fig:distance_dependence}b shows the model results for $V^0_{\mathrm{nlSSE}}$ (red), and the voltage measured at $H=H_{\mathrm{TA}}$ for $\delta=2$ (green) and $\delta=0.5$ (purple). While the voltage obtained from the model is approximately one order of magnitude lower than in experiments, the qualitative behaviour of the experimental data is reproduced. In particular, the modelled $d_{sc}$ approximately agrees with the experimentally observed distance. 

For $\delta=2$, the modelled voltage at $H_{\mathrm{TA}}$ is always enhanced with respect to $V^0_{\mathrm{nlSSE}}$ (for $d<d_{sc}$, $V(H_{\mathrm{TA}})<V^0_{\mathrm{nlSSE}}$ and for $d>d_{sc}$, $V(H_{\mathrm{TA}})>V^0_{\mathrm{nlSSE}}$). This is not consistent with the experiments as it leads to a sign change in $V_{\mathrm{TA}}$, which is defined as $V_{\mathrm{TA}}=V^0_{\mathrm{nlSSE}}-V(H_{\mathrm{TA}})$, as can be seen from Fig.~\ref{fig:distance_dependence}c. 

However, for $\delta=0.5$, $V(H_{\mathrm{TA}})$ is enhanced with respect to $V^0_{\mathrm{nlSSE}}$ for $d<d_{sc}$ but suppressed for $d>d_{sc}$. This results in a positive sign for $V_{\mathrm{TA}}$ over the full distance range, comparable to the experimental observations. The full magnetic field dependence obtained from the model can be found in the Supplementary Material (section F). As can be seen from the inset in Fig.~\ref{fig:distance_dependence}c, $\delta=0.5$ results in a decay of $V_{\mathrm{TA}}$ with distance which is comparable to the experimentally observed $V_{\mathrm{TA}}(d)$ (inset Fig.~\ref{fig:distance_dependence}a). We fitted the data for $V_{\mathrm{TA}}$ obtained from both the experiments and the simulations to $V_{\mathrm{TA}}(d)=A \exp{-d/\ell_{\mathrm{TA}}}$, where $A$ is the amplitude and $\ell_{\mathrm{TA}}$ the length scale over which $V_{\mathrm{TA}}$ decays. From the fits, we obtain $\ell_{\mathrm{TA}}^{\mathrm{exp}}=6.3\pm1.2$ $\mu$m and $\ell_{\mathrm{TA}}^{\mathrm{sim}}=10.6\pm0.1$ $\mu$m at room temperature, where we have fitted to the model results for $\delta=0.5$. From the simulations, we find that $\ell_{\mathrm{TA}}$ is influenced by the value used for $\delta$, where a smaller $\delta$ leads to a longer $\ell_{\mathrm{TA}}$. This could indicate that $\delta$ has to be increased slightly to obtain better agreement between $\ell_{\mathrm{TA}}^{\mathrm{exp}}$ and $\ell_{\mathrm{TA}}^{\mathrm{sim}}$. 

Therefore, in order to explain the observations, $0.5<\delta<1$, i.e. the relative enhancement due to magnon-phonon hybridization in $\sigma_m$ has to be larger than that of $\zeta$. $\ell_{\mathrm{TA}}^{\mathrm{exp}}$ is enhanced at low temperatures (see Supplementary Material (section B) for the distance dependence of $V_{\mathrm{TA}}$ at low temperatures). This could indicate that $\delta$ decreases with decreasing temperatures. For further discussion we refer to the Supplementary Material (section B).

The model results depend sensitively on $g_s$. A larger $g_s$ reduces the $d_{sc}$ observed in the model, so that our model no longer qualitatively fits the distance dependence of $V_{\mathrm{nlSSE}}$ obtained in experiments. As a consequence, the $\delta$ needed to model the resonant suppression of the signal at $H_{\mathrm{TA}}$ for long distances decreases further, which would imply that the enhancement in $\sigma_m$ is much stronger than that in $\zeta$. Such a strong enhancement in $\sigma_m$ should result in a clear magnon-polaron resonance in the electrically generated magnon spin signal, whereas we observed only a small effect here (see Fig.~\ref{fig:first_second_harmonic}a). This is an indication that our choice of reducing $g_s$ compared to our previous work is justified. 

\section{Discussion}
\label{sec:discussion}
We report resonant features in the nlSSE as a function of magnetic field, which we ascribe to the hybridization of magnons and acoustic phonons. They occur at magnetic fields that obey the ``touch'' condition at which the magnon frequency and group velocity agree with that of the TA and LA phonons. The signals are enhanced (peaks) for short injector-detector distances and high temperatures, but suppressed (dips) for long distances and/or low temperatures. The temperature dependence of the TA resonance differs from that of the low-field nlSSE voltage, indicating that different physical mechansims are involved (this in contrast to the local SSE configuration). The sign of the nlSSE signal corresponds to that of the signal in the hiSSE configuration for distances below the sign-change distance. In this regime the magnon-polaron feature causes signal enhancement, similar to the hiSSE configuration. For distances longer than the sign-change distance, the nlSSE signal is suppressed at the resonance magnetic field.

These results are consistent with a model in which transport is diffuse and carried by strongly coupled magnons and phonons\cite{Flebus2017} (magnon-polarons). Theory predicts an enhancement of all transport coefficients when the acoustic quality of the crystal is better than the magnetic one. Simulations show that the dip observed in the nlSSE is not caused by deteriorated acoustics, but by a competition between the thermally generated, SSE driven magnon current and the diffuse backflow magnon current which are both enhanced at the resonance. More experiments including thermal transport as well as an extension of the Boltzmann treatment presented in Ref.~\onlinecite{Flebus2017} to 2D geometries are necessary to fully come to grips with heat and spin transport in YIG. 

Additionally, we observed features in the electrically generated magnon spin signal at the resonance magnetic field. This is further evidence that not only the generation of magnons via the SSE, but additional transport parameters such as the magnon spin conductivity are affected by magnon-polarons.

The nonlocal measurement scheme provides an excellent platform to study magnon transport phenomena and opens up new avenues for studying the magnetoelastic coupling in magnetic insulators. Finally, these results are an important step towards a complete physical picture of magnon transport in magnetic insulators in its many aspects, which is crucial for developing efficient magnonic devices.

\begin{figure*}[h]
	\includegraphics[width=18.0cm]{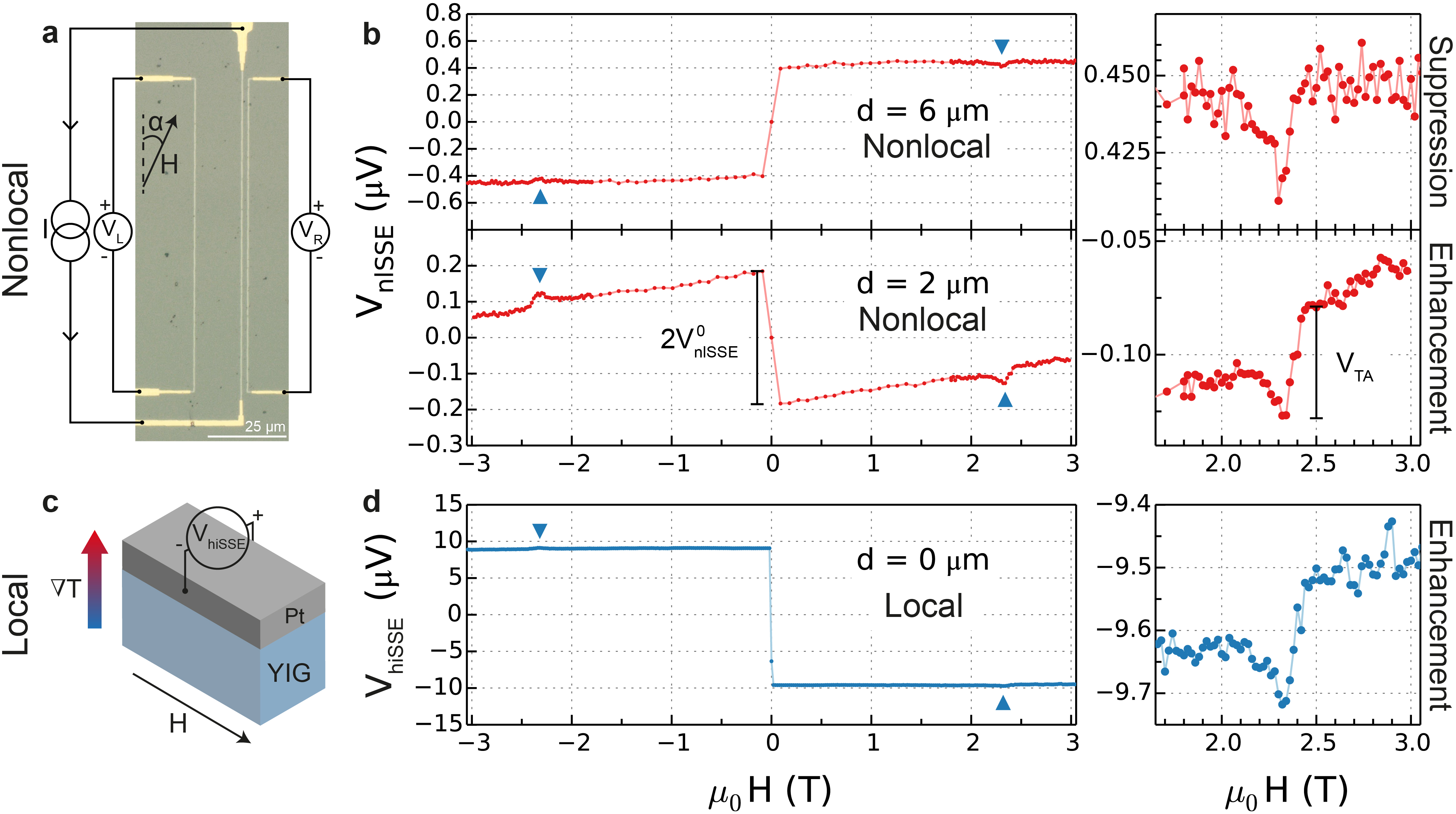}
	\caption{\textbf{Experimental geometries and main results.} Figure \textbf{a} is an image of a typical device, with schematic current and voltage connections. The three parallel lines are the Pt injector/detector strips, connected by Ti/Au contacts. $\alpha$ is the angle between the Pt strips and an applied magnetic field $H$ (in \textbf{b}-\textbf{d} $\alpha=90^{\circ}$). \textbf{b} The nonlocal spin Seebeck (nlSSE) voltage for an injector-detector distance $d=6$ $\mathrm{\mu}$m (top) and $d=2$ $\mathrm{\mu}$m (bottom) as a function of $\mu_0 H$. At $|\mu_0 H|=|\mu_0 H_{\mathrm{TA}}|\approx2.3$ T, a resonant structure is observed that we interpret in terms of magnon-polaron formation (indicated by blue triangles as a guide to the eye). The right column is a close-up of the anomalies for $H>0$. The results can be summarized by the voltages $V^0_{\mathrm{nlSSE}}$ and $V_{\mathrm{TA}}$ as indicated in the lower panels. \textbf{c} Schematic geometry of the local heater-induced hiSSE measurements. Here the temperature gradient $\nabla T$ is applied by external Peltier elements on the top and bottom of the sample. \textbf{d} The hiSSE voltage measured as a function of magnetic field. The close-up around the resonance field (right column) focusses on the magnon-polaron anomaly. All results were obtained at $T=200$ K. The results for $d=6$, $d=2$ and $d=0$ $\mu$m were obtained from sample S1, S2, S3, respectively (see Methods for sample details).}
	\label{fig:figure1}%
\end{figure*}

\begin{figure*}[h]
	\includegraphics[width=15.5cm]{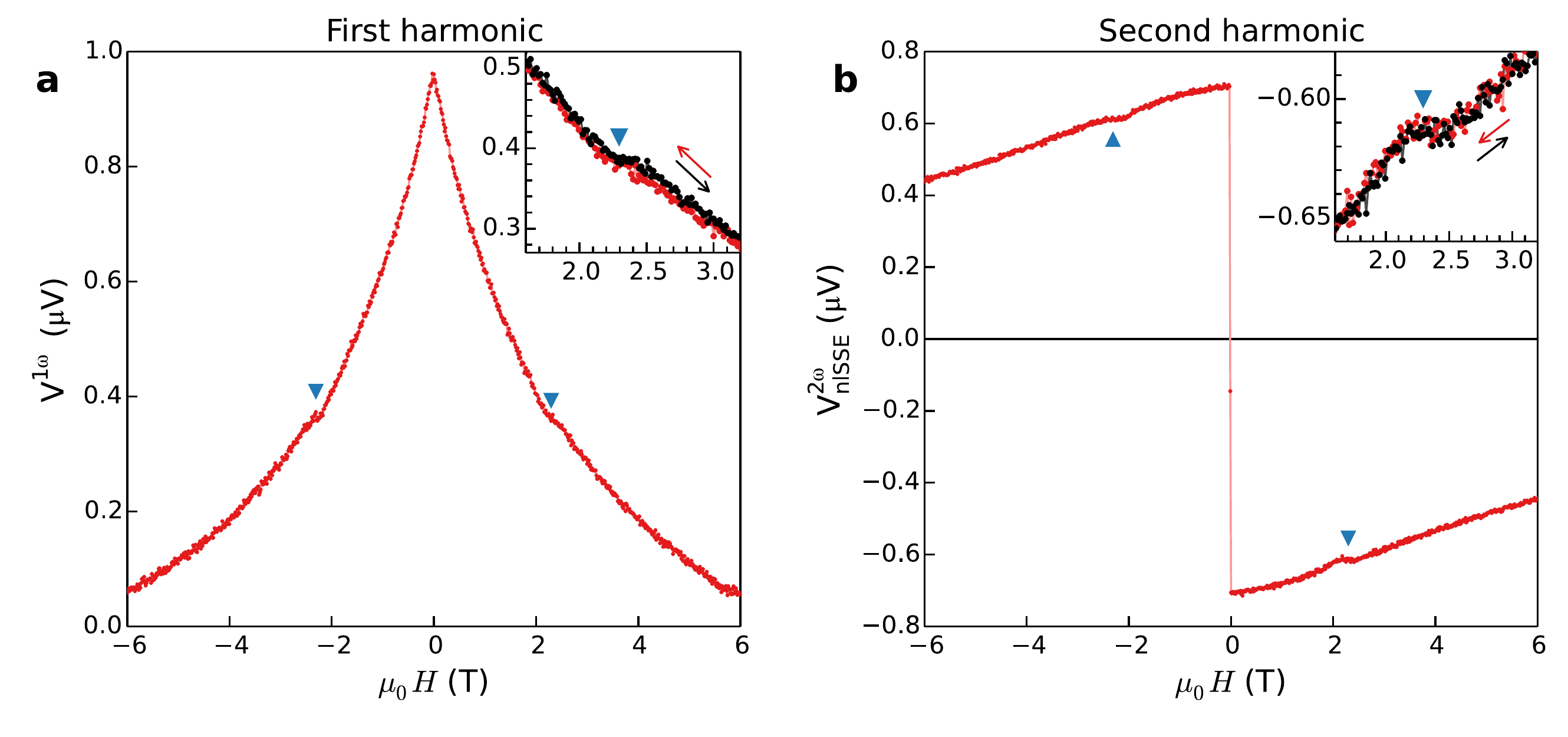}
	\caption{\textbf{Nonlocal voltage due to electrically and thermally generated magnons as a function of magnetic field.} Figure \textbf{a} shows the nonlocal voltage generated by magnons that are excited electrically (first harmonic response to an oscillating current in the injector contact). An anomaly is observed at $H=|H_{\mathrm{TA}}|$ (the field that satisfies the touching condition for magnons and transverse acoustic phonons). The inset shows a second set of data from the same sample, taken with a higher magnetic field resolution ($\mu_0\Delta H= 15$ mT), sweeping the magnetic field both in the forward (black) and backward (red) directions. Figure \textbf{b} shows the nlSSE voltage (second harmonic response) for the same device. $V_{\mathrm{nlSSE}}$ is suppressed at $H=|H_{\mathrm{TA}}|$. The inset shows the corresponding second harmonic data of the high resolution field sweep. The results were obtained on sample G1 (thickness 210 nm) with $d=3.5\,\mathrm{\mu m}$ and $I=150\,\mathrm{\mu A_{r.m.s.}}$, at room temperature. A constant background voltage $V_{\mathrm{bg}}=575$ nV was subtracted from the data in Fig. \textbf{a}.}  
	\label{fig:first_second_harmonic}%
\end{figure*}

\begin{figure*}[h]
	\includegraphics[width=18cm]{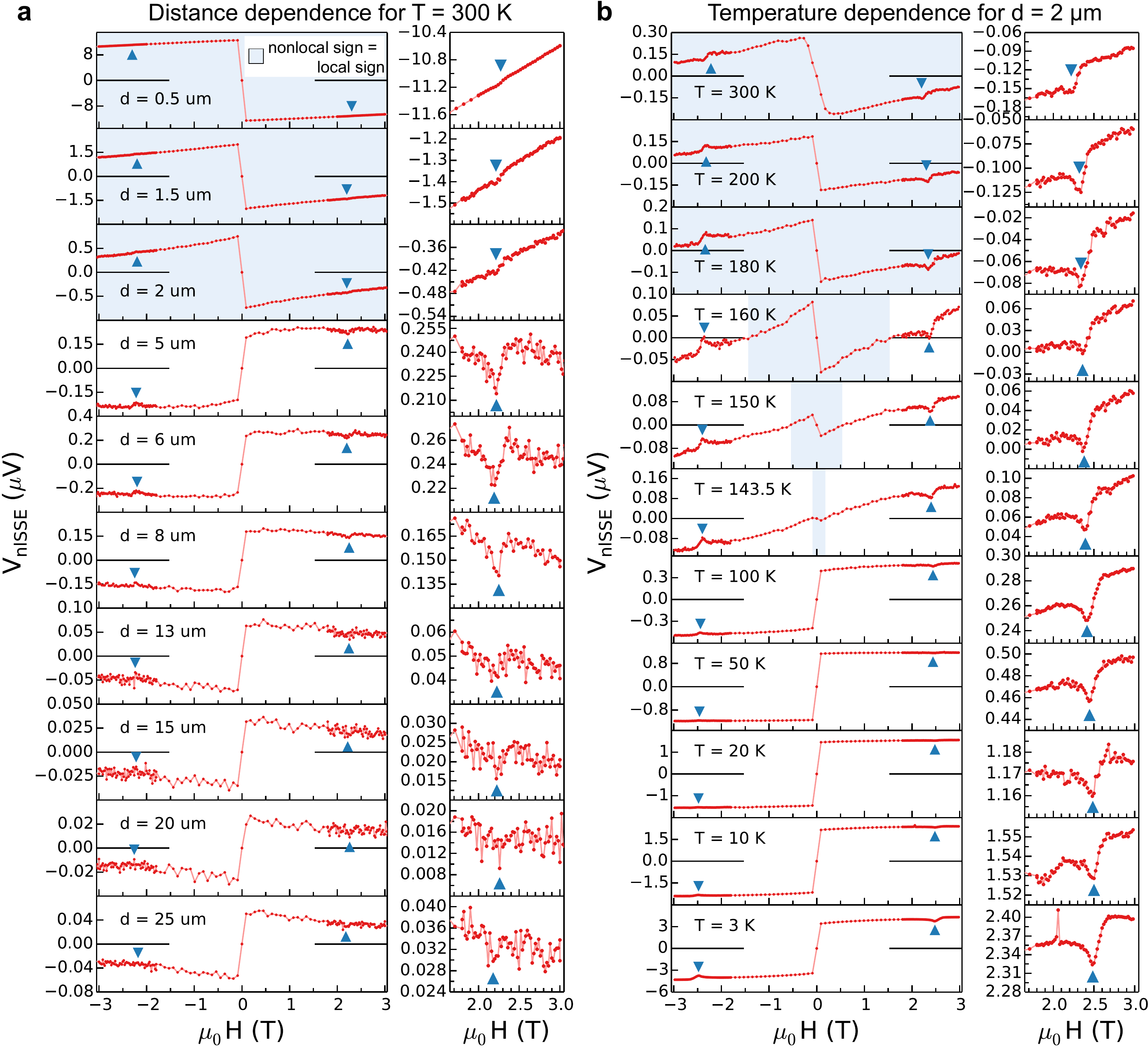}
	\caption{\textbf{$V_{\mathrm{nlSSE}}$ vs magnetic field as a function of distance and temperature.} Figure \textbf{a} is a plot of $V_{\mathrm{nlSSE}}$ vs $H$ for various injector-detector separations at $T=300$ K, while Figure \textbf{b} shows $V_{\mathrm{nlSSE}}$ vs $H$ for different temperatures and $d=2$ $\mu$m. The data in Figs. \textbf{a} and \textbf{b} are from sample S1 and S2, respectively. The  magnon-polaron resonance is indicated by the blue arrows. The blue shading in the graphs indicates the region in which the sign of the nlSSE signal agrees with that of the hiSSE. The right column in both \textbf{a} and \textbf{b} shows close-ups of the data around the positive resonance field (blue triangles). The data in the close-ups has been antisymmetrized with respect to $H$, i.e. $V=(V(+H)-V(-H))/2$. Fig.~\textbf{a} shows that when the contacts are close ($d\leq2$ $\mu$m), the magnon-polaron resonance enhances $V_{\mathrm{nlSSE}}$, while for long distances $V_{\mathrm{nlSSE}}$ is suppressed at the resonance magnetic field. For very large distances ($d\geq20$ $\mu$m), the resonance cannot be observed anymore. Similarly in Fig. \textbf{b}, for temperatures $T\geq180$ K, the magnon-polaron resonance enhances the nlSSE signal, while for lower temperatures the nlSSE signal is suppressed. The excitation current $I=100$ $\mu\mathrm{A_{r.m.s.}}$ for all measurements.}
	\label{fig:figure2}%
\end{figure*}

\begin{figure*}[h]
	\includegraphics[width=16.5cm]{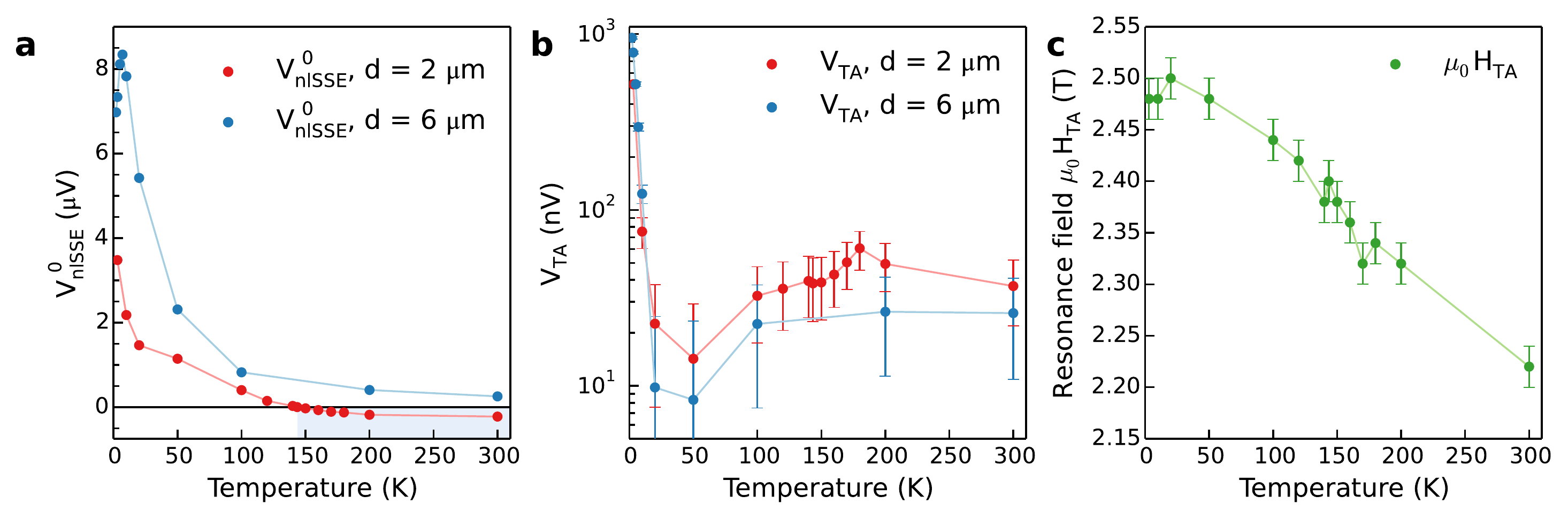}
	\caption{\textbf{Temperature dependence of $V^0_{\mathrm{nlSSE}}$, $V_{\mathrm{TA}}$ and $H_{\mathrm{TA}}$.} \textbf{a} displays the temperature dependence of the low-field $V^0_{\mathrm{nlSSE}}$, for $d=2$ $\mathrm{\mu}$m and $d=6$ $\mathrm{\mu}$m. For $2$ $\mathrm{\mu m}$, the signal changes sign around $T=143$ K. The blue shading in the graph indicates the regime in which the sign agrees with that of the hiSSE. The temperature dependence of the magnon-polaron resonance $V_{\mathrm{TA}}$ is shown in Figure \textbf{b}. Here, no sign change but a minimum around $T=50$ K is observed, which is absent in Figure \textbf{a}. Figure \textbf{c} shows the temperature dependence of the resonance field $H_{\mathrm{TA}}$. Error bars in \textbf{b} and \textbf{c} reflect the peak-to-peak noise in the data used to extract $V_{\mathrm{TA}}$ and the step size in the magnetic field scans ($\mu_0\Delta H=20$ mT), respectively.}
	\label{fig:figure3}%
\end{figure*}

\begin{figure*}[h]
	\includegraphics[width=16.5cm]{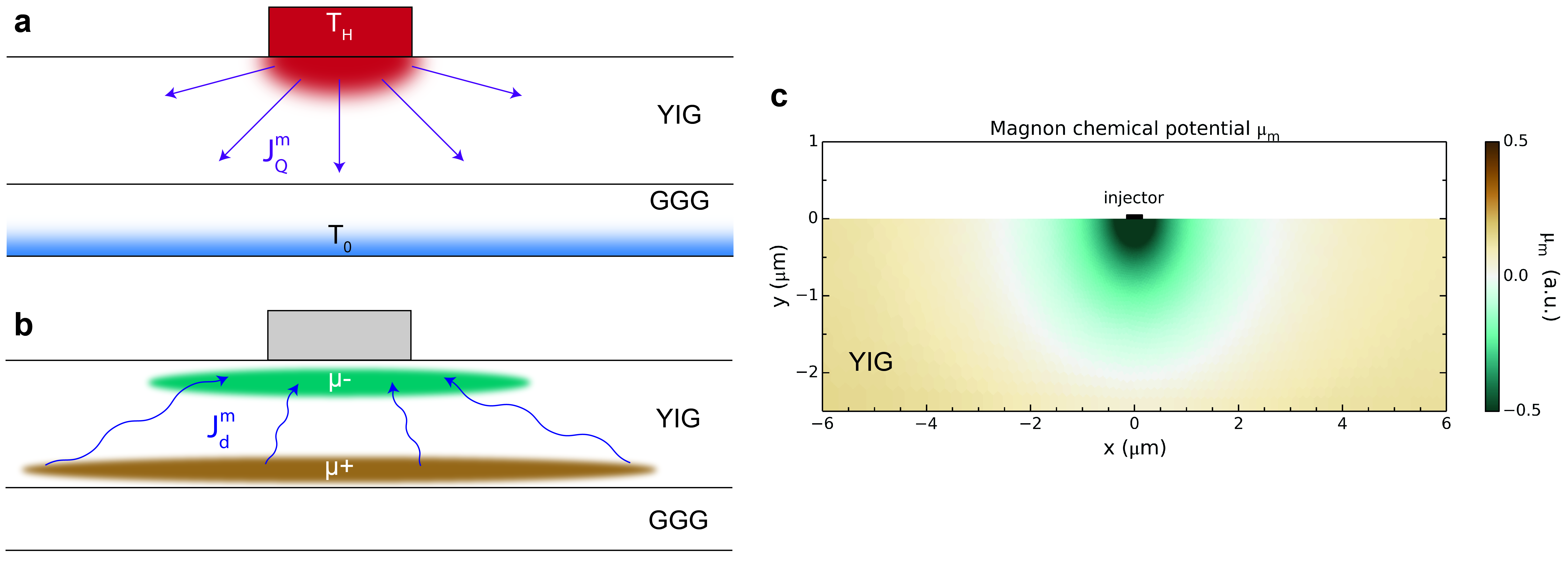}
	\caption{\textbf{Physical concepts underlying the nlSSE signal and simulated magnon chemical potential profile.} Figure \textbf{a} sketches the effects of Joule heating in the injector, heating it up to temperature $T_H$, which leads to a thermal gradient in the YIG. The bulk SSE generates a magnon current $J^m_Q$ antiparallel to the local temperature gradient, spreading into the film away from the contact. When the spin conductance of the contact is sufficiently small, this leads to a depletion of magnons below the injector, indicated in Figure \textbf{b} as $\upmu^-$. When the magnons are reflected at the GGG interface, $J^m_Q$ accumulates magnons at the YIG\textbar{}GGG interface, shown in Figure \textbf{b} as $\upmu^+$. The chemical potential gradient induces a backward and sideward diffuse magnon current $J^m_d$. Both processes in Figure \textbf{a} and \textbf{b} are included in the finite element model (FEM). Its results are plotted in Figure \textbf{c} in terms of a typical magnon chemical potential profile. $\mu_m$ changes sign at some distance from the injector, also at the YIG surface, where it can be detected by a second contact. The magnon-polaron resonance enhances both the spin Seebeck coefficient $\zeta$ and the magnon spin conductivity $\sigma_m$. The increased backflow of magnons to the injector causes a suppression of the nonlocal signal at long distances (see Figure 6).}
	\label{fig:figure5}%
\end{figure*}

\begin{figure*}[h]
	\includegraphics[width=18.0cm]{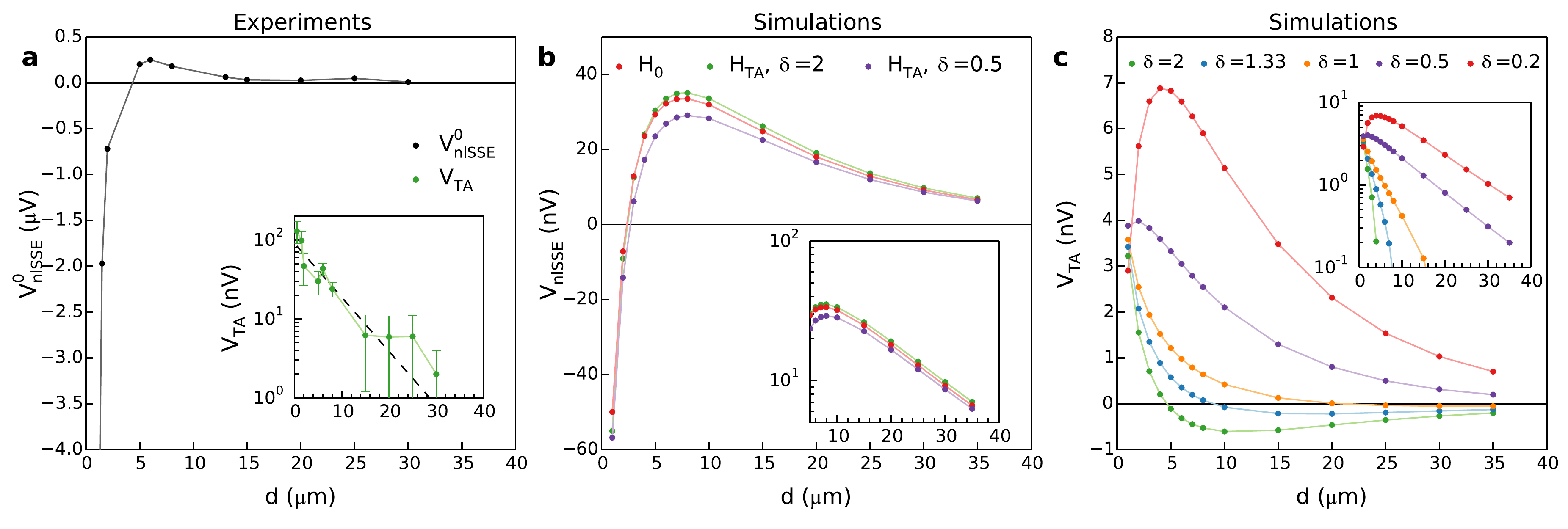}
	\caption{\textbf{Comparison of the experimental and simulated $V^0_{\mathrm{nlSSE}}$ and $V_{\mathrm{TA}}$.} Figure \textbf{a} shows the distance dependence of $V^0_{\mathrm{nlSSE}}$ and $V_{\mathrm{TA}}$ (inset) measured at room temperature. The dashed line in the inset is an exponential fit to the data. $V^0_{\mathrm{nlSSE}}$ changes sign around $d=4\,\mathrm{\mu m}$, while $V_{\mathrm{TA}}$ remains positive. Figure \textbf{b} is a plot of the calculated distance dependence of $V^0_{\mathrm{nlSSE}}$ at zero magnetic field (red) and at the resonant field for $\delta=2$ (green) and $\delta=0.5$ (purple). Here $\delta$ is a parameter that measures the relative enhancement of the spin Seebeck coefficient compared to the magnon spin conductivity, as explained in the main text. The inset shows the signal decay at long distances on a logarithmic scale. Figure \textbf{c} shows the modelled distance dependence of $V_{\mathrm{TA}}$ for various values of $\delta$ on a linear scale (inset for logarithmic scale). $\delta=0.5$ results in a positive sign for $V_{\mathrm{TA}}$ over the full distance range with a slope that roughly agrees with experiments (cf. insets of Figure \textbf{a} and \textbf{c}). Reducing $\delta$ further leads to a more gradual slope for $V_{\mathrm{TA}}$. In the simulations, the SSE enhancement is $f_{\zeta}=1.09$, while $f_{\sigma}$ is varied with $\delta$.}
	\label{fig:distance_dependence}%
\end{figure*}

\section{Methods}
\label{sec:methods}
\textbf{Sample fabrication.} The YIG films used in this study were all grown on gadolinium gallium garnet (GGG) substrates by liquid phase epitaxy (LPE) in the [111] direction. The samples from the Sendai group have a thickness of 2.5 $\upmu$m, the samples used in Groningen are 210 nm thick. The Sendai samples were grown in-house, whereas the Groningen samples were obtained commercially from Matesy GmbH. In Sendai, five batches of devices where fabricated from the same YIG wafer (S1 to S4). The fabrication method and platinum strip geometry are the same for all batches, but they were not fabricated at the same time, which might lead to variations in for instance the interface quality from batch to batch. In Groningen, two batches of devices were investigated (G1 and G2). The nonlocal devices fabricated in Groningen are defined in three lithography steps: the first step was used to define Ti/Au markers on top of the YIG film via e-beam evaporation, used to align the subsequent steps. In the second step, Pt injector and detector strips were deposited using magnetron sputtering in an Ar$^+$ plasma. In the final step, Ti/Au contacts were deposited by e-beam evaporation. Prior to the contact deposition, a brief Ar$^+$ ion beam etching step was performed to remove any polymer residues from the Pt strip contact areas to ensure optimal electrical contact to the devices. The nonlocal devices fabricated in Sendai were defined in a single lithography step. Two parallel Pt strips and contact pads were patterned using  e-beam lithography followed by a lift-off process, in which 10-nm-thick Pt was deposited using magnetron sputtering in an Ar$^{+}$ plasma.

\textbf{Measurements.} Electrical measurements were carried out in Groningen and in Sendai, using a current-biased lock-in detection scheme. A low frequency ac current of angular frequency $\omega$ (typical frequencies are $\omega/(2\pi)<20$ Hz, and the typical amplitude is $I=100$ $\mathrm{\mu A_{rms}}$) is sent through the injector strip, and the voltage on the detector strip is measured at both the frequencies $\omega$ (the first harmonic response) and $2\omega$ (the second harmonic response). This allows us to separate processes that are linear in the current, which govern the first harmonic response, from processes that are quadratic in the current which are measured in the second harmonic response\cite{Bakker2010,Vlietstra2014, Cornelissen2015}. 

The measurements in Sendai were carried out in a Quantum Design Physical Properties Measurement System (PPMS), using a superconducting solenoid to apply the external magnetic field (field range up to $\mu_0H=\pm10.5$ T). The measurements in Groningen were carried out in a cryostat equipped with a Cryogenics Limited variable temperature insert (VTI) and superconducting solenoid (magnetic field range up to $\mu_0H=\pm7.5$ T). Electronic measurements in Groningen are carried out using a home built current source and voltage pre-amplifier (gain $10^4$) module galvanically isolated from the rest of the measurement electronics, resulting in a noise level of approximately $3$ n$\mathrm{V_{r.m.s.}}$ at the output of the lockin amplifier for a time constant of $\tau=3$ s and a filter slope of $24$ dB/octave. The electronic measurements in Sendai were carried out by means of an ac and dc current source (Keithley model 6221) and a lockin amplifier using a time constant of $\tau=1$ s and a filter slope of $24$ dB/octave. The data shown in Figure~\ref{fig:figure1}b and Figure~\ref{fig:figure2} is the asymmetric part of the measured voltage with respect to the magnetic field. The antisymmetrization procedure includes both the forward and backward magnetic field sweep, and the voltage shown in the figures is given by $V_{\mathrm{H+}}=(V_{\mathrm{backward}}(H)-V_{\mathrm{backward}}(-H))/2$ and $V_{\mathrm{H-}}=(V_{\mathrm{forward}}(H)-V_{\mathrm{forward}}(-H))/2$, where $V_{\mathrm{H+}}$ is the voltage at postive magnetic field values and $V_{\mathrm{H-}}$ that at negative magnetic field values.

\textbf{Simulations.} The two-dimensional finite element model is implemented in COMSOL MultiPhysics (v4.4). The linear response relation of heat and spin transport in the bulk of a magnetic insulator reads
\begin{equation}%
\begin{pmatrix}
\frac{2e}{\hbar}\mathbf{j}_{m}\\
\mathbf{j}_{\mathrm{Q}}%
\end{pmatrix}
=-%
\begin{pmatrix}
\sigma_{m} & \zeta/T\\
\hbar \zeta/2e & \kappa%
\end{pmatrix}%
\begin{pmatrix}
\boldsymbol{\nabla}\mu_{m}\\
\boldsymbol{\nabla}T%
\end{pmatrix}
,\label{eq:linearresponsespinheatmagnet}%
\end{equation} where $\mathbf{j}_m$ is the magnon spin current, $\mathbf{j}_{\mathrm{Q}}$ the total (magnon and phonon) heat current, $\mu_m$ the magnon chemical potential, $T$ the temperature (assumed to be the same for magnons and phonons by efficient thermalization), $\sigma_m$ the magnon spin conductivity, $\kappa$ the total (magnon and phonon) heat conductivity and $\zeta$ the spin Seebeck coefficient. We disregard temperature differences arising from the Kapitza resistances at the Pt\textbar{}YIG or YIG\textbar{}GGG interfaces. $-e$ is the electron charge and $\hbar$ the reduced Planck constant. The diffusion equations for spin and heat read
\begin{align}
\nabla^2\mu_m &=\frac{\mu_m}{\lambda_m^2}, \label{eq:spinrelaxation}\\
\nabla^2T &= \frac{j_c^2}{\kappa\sigma}, \label{eq:sourcet}
\end{align}
where $j_c$ is the charge current density in the injector contact, $\sigma$ and $\kappa$ the electrical and thermal conductivity and $\lambda_m$ the magnon spin diffusion length. Eq.~(\ref{eq:sourcet}) represents the Joule heating in the injector that drives the SSE.

In the simulations, $t_{\mathrm{YIG}}=2.5$ $\mathrm{\mu}$m and $w_{\mathrm{YIG}}=500$ $\mathrm{\mu}$m are the thickness and width of the YIG film, on top of a GGG substrate that is $500$ $\mathrm{\mu}$m thick. $w_{\mathrm{YIG}}$ is much larger than $\lambda_m$ and finite size effects are absent. The injector has a thickness of $t_{\mathrm{Pt}}=10$ nm and a width of $w_{\mathrm{Pt}}=300$ nm. The spin and heat currents normal to the YIG\textbar{}vacuum, Pt\textbar{}vacuum and GGG\textbar{}vacuum interfaces vanish. At the bottom of the GGG substrate the boundary condition $T=T_0$ is used, i.e. the bottom of the sample is taken to be thermally anchored to the sample probe. Furthermore, a spin current is not allowed to flow into the GGG. The spin current across the Pt\textbar{}YIG interface is given by ${j_m^\mathrm{int}=g_s\left(\mu_s-\mu_m\right)}$, where $g_s$ is the effective spin conductance of the interface, $\mu_s$ is the spin accumulation on the metal side of the interface and $\mu_m$ is the magnon chemical potential on the YIG side of the interface. The nonlocal voltage is then found by calculating the average spin current density $\langle j_s \rangle$ flowing in the detector, which is then converted to non-local voltage using $V_{\rm nlSSE}=\theta_{\mathrm{SH}}L \langle j_s \rangle /\sigma$, where $\theta_{\mathrm{SH}}$ is the spin Hall angle in platinum and $L$ is the length of the detector strip. The spin current in the platinum contact relaxes over the characteristic spin relaxation length $\lambda_s$.

The parameters we use for platinum in the model are $\theta_{\mathrm{SH}}=0.11$, $\sigma=1.9\times10^6$ S/m, $\lambda_s=1.5$ nm and $\kappa=26$ W/(m K). For YIG, we use $\sigma_m=3.7\times10^5$ S/m, $\lambda_m=9.4$ $\mathrm{\mu}$m which was obtained in our previous work\cite{Cornelissen2016b}. Furthermore, we use $\kappa=7$ W/(m K), based on YIG thermal conductivity data from Ref.~\onlinecite{PhysRevB.90.064421}. For the bulk spin Seebeck coefficient at zero field we use $\zeta^0=500$ A/m, based on our previous work in which we gave an estimate for $\zeta$ at room temperature\cite{Shan}. For GGG, the spin conductivity and spin Seebeck coefficient are set to zero. For the GGG thermal conductivity we use $\kappa=9$ W/(m K), based on data from Refs.~\onlinecite{Daudin1982,Slack1971}. Finally, for the effective spin conductance of the interface we use $g_s=3.4\times10^{11}$ S/m$^2$. We note that this is roughly a factor 30 smaller than in our earlier work\cite{Cornelissen2016b}. This variation of the interface transparency in different experiments indicates the presence of physical processes that are not taken into account in the modeling.

\section{Acknowledgements}
\label{sec:acknowledgements}
We thank H. M. de Roosz, J.G. Holstein, H. Adema and T.J. Schouten for technical assistance and R.A. Duine, B. Flebus and K. Shen for discussions. This work is part of the research program of the Netherlands Organization for Scientific Research (NWO) and supported by NanoLab NL, EU FP7 ICT Grant No. 612759 InSpin, the Zernike Institute for Advanced Materials, Grant-in-Aid for Scientific Research on Innovative Area "Nano Spin Conversion Science" (Nos. JP26103005 and JP26103006), Grant-in-Aid for Scientific Research (A) (No. JP25247056) and (S) (No. JP25220910) from JSPS KAKENHI, Japan, and ERATO "Spin Quantum Rectification Project" (No. JPMJER1402) from JST, Japan. Further support by the DFG priority program Spin Caloric Transport (SPP 1538, KU3271/1-1) is gratefully acknowledged. K.O. acknowledges support from GP-Spin at Tohoku University. T.Ki. is supported by JSPS through a research fellowship for young scientists (No. JP15J08026).

\section{Author contributions}
\label{sec:authorcontributions}
{B.J.v.W., L.J.C., T.Ki. and E.S. conceived the experiments. Z.Q. fabricated the Sendai YIG films. K.O. and L.J.C. fabricated the nonlocal devices in Sendai and Groningen, respectively. K.O. and L.J.C. performed the experiments. T.Ki. supervised the experiments in Sendai. K.O., L.J.C., T.Ki., T.Ku., G.E.W.B. and E.S. analyzed and interpreted the data. L.J.C. performed the numerical modelling. L.J.C., T.Ku. and  G.E.W.B. interpreted the modelling results. L.J.C. wrote the paper, with the help of all co-authors.
%\appendix
%\section{Measurements of electrically generated magnons}
%\section{Local current-driven spin Seebeck effect}
%\section{Comparison between samples from Sendai and Groningen}

%\bibliographystyle{plain}
\bibliography{bibliography}

%merlin.mbs apsrev4-1.bst 2010-07-25 4.21a (PWD, AO, DPC) hacked
%Control: key (0)
%Control: author (8) initials jnrlst
%Control: editor formatted (1) identically to author
%Control: production of article title (-1) disabled
%Control: page (0) single
%Control: year (1) truncated
%Control: production of eprint (0) enabled
\begin{thebibliography}{47}%
\makeatletter
\providecommand \@ifxundefined [1]{%
 \@ifx{#1\undefined}
}%
\providecommand \@ifnum [1]{%
 \ifnum #1\expandafter \@firstoftwo
 \else \expandafter \@secondoftwo
 \fi
}%
\providecommand \@ifx [1]{%
 \ifx #1\expandafter \@firstoftwo
 \else \expandafter \@secondoftwo
 \fi
}%
\providecommand \natexlab [1]{#1}%
\providecommand \enquote  [1]{``#1''}%
\providecommand \bibnamefont  [1]{#1}%
\providecommand \bibfnamefont [1]{#1}%
\providecommand \citenamefont [1]{#1}%
\providecommand \href@noop [0]{\@secondoftwo}%
\providecommand \href [0]{\begingroup \@sanitize@url \@href}%
\providecommand \@href[1]{\@@startlink{#1}\@@href}%
\providecommand \@@href[1]{\endgroup#1\@@endlink}%
\providecommand \@sanitize@url [0]{\catcode `\\12\catcode `\$12\catcode
  `\&12\catcode `\#12\catcode `\^12\catcode `\_12\catcode `\%12\relax}%
\providecommand \@@startlink[1]{}%
\providecommand \@@endlink[0]{}%
\providecommand \url  [0]{\begingroup\@sanitize@url \@url }%
\providecommand \@url [1]{\endgroup\@href {#1}{\urlprefix }}%
\providecommand \urlprefix  [0]{URL }%
\providecommand \Eprint [0]{\href }%
\providecommand \doibase [0]{http://dx.doi.org/}%
\providecommand \selectlanguage [0]{\@gobble}%
\providecommand \bibinfo  [0]{\@secondoftwo}%
\providecommand \bibfield  [0]{\@secondoftwo}%
\providecommand \translation [1]{[#1]}%
\providecommand \BibitemOpen [0]{}%
\providecommand \bibitemStop [0]{}%
\providecommand \bibitemNoStop [0]{.\EOS\space}%
\providecommand \EOS [0]{\spacefactor3000\relax}%
\providecommand \BibitemShut  [1]{\csname bibitem#1\endcsname}%
\let\auto@bib@innerbib\@empty
%</preamble>
\bibitem [{\citenamefont {Kittel}(1958)}]{Kittel1958}%
  \BibitemOpen
  \bibfield  {author} {\bibinfo {author} {\bibfnamefont {C.}~\bibnamefont
  {Kittel}},\ }\href {\doibase 10.1103/PhysRev.110.836} {\bibfield  {journal}
  {\bibinfo  {journal} {Physical Review}\ }\textbf {\bibinfo {volume} {110}},\
  \bibinfo {pages} {836} (\bibinfo {year} {1958})}\BibitemShut {NoStop}%
\bibitem [{\citenamefont {Eshbach}(1963)}]{Eshbach1963}%
  \BibitemOpen
  \bibfield  {author} {\bibinfo {author} {\bibfnamefont {J.~R.}\ \bibnamefont
  {Eshbach}},\ }\href {\doibase 10.1063/1.1729481} {\bibfield  {journal}
  {\bibinfo  {journal} {Journal of Applied Physics}\ }\textbf {\bibinfo
  {volume} {34}},\ \bibinfo {pages} {1298} (\bibinfo {year}
  {1963})}\BibitemShut {NoStop}%
\bibitem [{\citenamefont {Schl{\"{o}}mann}\ and\ \citenamefont
  {Joseph}(1964)}]{Schlomann1964}%
  \BibitemOpen
  \bibfield  {author} {\bibinfo {author} {\bibfnamefont {E.}~\bibnamefont
  {Schl{\"{o}}mann}}\ and\ \bibinfo {author} {\bibfnamefont {R.~I.}\
  \bibnamefont {Joseph}},\ }\href {\doibase 10.1063/1.1702867} {\bibfield
  {journal} {\bibinfo  {journal} {Journal of Applied Physics}\ }\textbf
  {\bibinfo {volume} {35}},\ \bibinfo {pages} {2382} (\bibinfo {year}
  {1964})}\BibitemShut {NoStop}%
\bibitem [{\citenamefont {R{\"{u}}ckriegel}\ \emph {et~al.}(2014)\citenamefont
  {R{\"{u}}ckriegel}, \citenamefont {Kopietz}, \citenamefont {Bozhko},
  \citenamefont {Serga},\ and\ \citenamefont
  {Hillebrands}}]{PhysRevB.89.184413}%
  \BibitemOpen
  \bibfield  {author} {\bibinfo {author} {\bibfnamefont {A.}~\bibnamefont
  {R{\"{u}}ckriegel}}, \bibinfo {author} {\bibfnamefont {P.}~\bibnamefont
  {Kopietz}}, \bibinfo {author} {\bibfnamefont {D.~A.}\ \bibnamefont {Bozhko}},
  \bibinfo {author} {\bibfnamefont {A.~A.}\ \bibnamefont {Serga}}, \ and\
  \bibinfo {author} {\bibfnamefont {B.}~\bibnamefont {Hillebrands}},\ }\href
  {\doibase 10.1103/PhysRevB.89.184413} {\bibfield  {journal} {\bibinfo
  {journal} {Physical Review B}\ }\textbf {\bibinfo {volume} {89}},\ \bibinfo
  {pages} {184413} (\bibinfo {year} {2014})}\BibitemShut {NoStop}%
\bibitem [{\citenamefont {Ogawa}\ \emph {et~al.}(2015)\citenamefont {Ogawa},
  \citenamefont {Koshibae}, \citenamefont {Beekman}, \citenamefont {Nagaosa},
  \citenamefont {Kubota}, \citenamefont {Kawasaki},\ and\ \citenamefont
  {Tokura}}]{Ogawa2015}%
  \BibitemOpen
  \bibfield  {author} {\bibinfo {author} {\bibfnamefont {N.}~\bibnamefont
  {Ogawa}}, \bibinfo {author} {\bibfnamefont {W.}~\bibnamefont {Koshibae}},
  \bibinfo {author} {\bibfnamefont {A.~J.}\ \bibnamefont {Beekman}}, \bibinfo
  {author} {\bibfnamefont {N.}~\bibnamefont {Nagaosa}}, \bibinfo {author}
  {\bibfnamefont {M.}~\bibnamefont {Kubota}}, \bibinfo {author} {\bibfnamefont
  {M.}~\bibnamefont {Kawasaki}}, \ and\ \bibinfo {author} {\bibfnamefont
  {Y.}~\bibnamefont {Tokura}},\ }\href {\doibase 10.1073/pnas.1504064112}
  {\bibfield  {journal} {\bibinfo  {journal} {Proceedings of the National
  Academy of Sciences}\ }\textbf {\bibinfo {volume} {112}},\ \bibinfo {pages}
  {8977} (\bibinfo {year} {2015})}\BibitemShut {NoStop}%
\bibitem [{\citenamefont {Kamra}\ \emph {et~al.}(2015)\citenamefont {Kamra},
  \citenamefont {Keshtgar}, \citenamefont {Yan},\ and\ \citenamefont
  {Bauer}}]{Kamra2015}%
  \BibitemOpen
  \bibfield  {author} {\bibinfo {author} {\bibfnamefont {A.}~\bibnamefont
  {Kamra}}, \bibinfo {author} {\bibfnamefont {H.}~\bibnamefont {Keshtgar}},
  \bibinfo {author} {\bibfnamefont {P.}~\bibnamefont {Yan}}, \ and\ \bibinfo
  {author} {\bibfnamefont {G.~E.~W.}\ \bibnamefont {Bauer}},\ }\href {\doibase
  10.1103/PhysRevB.91.104409} {\bibfield  {journal} {\bibinfo  {journal}
  {Physical Review B}\ }\textbf {\bibinfo {volume} {91}},\ \bibinfo {pages}
  {104409} (\bibinfo {year} {2015})}\BibitemShut {NoStop}%
\bibitem [{\citenamefont {Shen}\ and\ \citenamefont {Bauer}(2015)}]{Shen2015}%
  \BibitemOpen
  \bibfield  {author} {\bibinfo {author} {\bibfnamefont {K.}~\bibnamefont
  {Shen}}\ and\ \bibinfo {author} {\bibfnamefont {G.~E.~W.}\ \bibnamefont
  {Bauer}},\ }\href {\doibase 10.1103/PhysRevLett.115.197201} {\bibfield
  {journal} {\bibinfo  {journal} {Physical Review Letters}\ }\textbf {\bibinfo
  {volume} {115}},\ \bibinfo {pages} {197201} (\bibinfo {year}
  {2015})}\BibitemShut {NoStop}%
\bibitem [{\citenamefont {Guerreiro}\ and\ \citenamefont
  {Rezende}(2015)}]{Guerreiro2015}%
  \BibitemOpen
  \bibfield  {author} {\bibinfo {author} {\bibfnamefont {S.~C.}\ \bibnamefont
  {Guerreiro}}\ and\ \bibinfo {author} {\bibfnamefont {S.~M.}\ \bibnamefont
  {Rezende}},\ }\href {\doibase 10.1103/PhysRevB.92.214437} {\bibfield
  {journal} {\bibinfo  {journal} {Physical Review B}\ }\textbf {\bibinfo
  {volume} {92}},\ \bibinfo {pages} {214437} (\bibinfo {year}
  {2015})}\BibitemShut {NoStop}%
\bibitem [{\citenamefont {Kikkawa}\ \emph
  {et~al.}(2016{\natexlab{a}})\citenamefont {Kikkawa}, \citenamefont {Shen},
  \citenamefont {Flebus}, \citenamefont {Duine}, \citenamefont {Uchida},
  \citenamefont {Qiu}, \citenamefont {Bauer},\ and\ \citenamefont
  {Saitoh}}]{Kikkawa2016}%
  \BibitemOpen
  \bibfield  {author} {\bibinfo {author} {\bibfnamefont {T.}~\bibnamefont
  {Kikkawa}}, \bibinfo {author} {\bibfnamefont {K.}~\bibnamefont {Shen}},
  \bibinfo {author} {\bibfnamefont {B.}~\bibnamefont {Flebus}}, \bibinfo
  {author} {\bibfnamefont {R.~A.}\ \bibnamefont {Duine}}, \bibinfo {author}
  {\bibfnamefont {K.-i.}\ \bibnamefont {Uchida}}, \bibinfo {author}
  {\bibfnamefont {Z.}~\bibnamefont {Qiu}}, \bibinfo {author} {\bibfnamefont
  {G.~E.~W.}\ \bibnamefont {Bauer}}, \ and\ \bibinfo {author} {\bibfnamefont
  {E.}~\bibnamefont {Saitoh}},\ }\href {\doibase
  10.1103/PhysRevLett.117.207203} {\bibfield  {journal} {\bibinfo  {journal}
  {Physical Review Letters}\ }\textbf {\bibinfo {volume} {117}},\ \bibinfo
  {pages} {207203} (\bibinfo {year} {2016}{\natexlab{a}})}\BibitemShut
  {NoStop}%
\bibitem [{\citenamefont {Flebus}\ \emph {et~al.}(2017)\citenamefont {Flebus},
  \citenamefont {Shen}, \citenamefont {Kikkawa}, \citenamefont {Uchida},
  \citenamefont {Qiu}, \citenamefont {Saitoh}, \citenamefont {Duine},\ and\
  \citenamefont {Bauer}}]{Flebus2017}%
  \BibitemOpen
  \bibfield  {author} {\bibinfo {author} {\bibfnamefont {B.}~\bibnamefont
  {Flebus}}, \bibinfo {author} {\bibfnamefont {K.}~\bibnamefont {Shen}},
  \bibinfo {author} {\bibfnamefont {T.}~\bibnamefont {Kikkawa}}, \bibinfo
  {author} {\bibfnamefont {K.-i.}\ \bibnamefont {Uchida}}, \bibinfo {author}
  {\bibfnamefont {Z.}~\bibnamefont {Qiu}}, \bibinfo {author} {\bibfnamefont
  {E.}~\bibnamefont {Saitoh}}, \bibinfo {author} {\bibfnamefont {R.~A.}\
  \bibnamefont {Duine}}, \ and\ \bibinfo {author} {\bibfnamefont {G.~E.~W.}\
  \bibnamefont {Bauer}},\ }\href {\doibase 10.1103/PhysRevB.95.144420}
  {\bibfield  {journal} {\bibinfo  {journal} {Physical Review B}\ }\textbf
  {\bibinfo {volume} {95}},\ \bibinfo {pages} {144420} (\bibinfo {year}
  {2017})}\BibitemShut {NoStop}%
\bibitem [{\citenamefont {Flipse}\ \emph {et~al.}(2014)\citenamefont {Flipse},
  \citenamefont {Dejene}, \citenamefont {Wagenaar}, \citenamefont {Bauer},
  \citenamefont {Youssef},\ and\ \citenamefont {van Wees}}]{Flipse2014}%
  \BibitemOpen
  \bibfield  {author} {\bibinfo {author} {\bibfnamefont {J.}~\bibnamefont
  {Flipse}}, \bibinfo {author} {\bibfnamefont {F.~K.}\ \bibnamefont {Dejene}},
  \bibinfo {author} {\bibfnamefont {D.}~\bibnamefont {Wagenaar}}, \bibinfo
  {author} {\bibfnamefont {G.~E.~W.}\ \bibnamefont {Bauer}}, \bibinfo {author}
  {\bibfnamefont {J.~B.}\ \bibnamefont {Youssef}}, \ and\ \bibinfo {author}
  {\bibfnamefont {B.~J.}\ \bibnamefont {van Wees}},\ }\href {\doibase
  10.1103/PhysRevLett.113.027601} {\bibfield  {journal} {\bibinfo  {journal}
  {Physical Review Letters}\ }\textbf {\bibinfo {volume} {113}},\ \bibinfo
  {pages} {027601} (\bibinfo {year} {2014})}\BibitemShut {NoStop}%
\bibitem [{\citenamefont {Agrawal}\ \emph {et~al.}(2013)\citenamefont
  {Agrawal}, \citenamefont {Vasyuchka}, \citenamefont {Serga}, \citenamefont
  {Karenowska}, \citenamefont {Melkov},\ and\ \citenamefont
  {Hillebrands}}]{Agrawal2013}%
  \BibitemOpen
  \bibfield  {author} {\bibinfo {author} {\bibfnamefont {M.}~\bibnamefont
  {Agrawal}}, \bibinfo {author} {\bibfnamefont {V.~I.}\ \bibnamefont
  {Vasyuchka}}, \bibinfo {author} {\bibfnamefont {A.~A.}\ \bibnamefont
  {Serga}}, \bibinfo {author} {\bibfnamefont {A.~D.}\ \bibnamefont
  {Karenowska}}, \bibinfo {author} {\bibfnamefont {G.~A.}\ \bibnamefont
  {Melkov}}, \ and\ \bibinfo {author} {\bibfnamefont {B.}~\bibnamefont
  {Hillebrands}},\ }\href {\doibase 10.1103/PhysRevLett.111.107204} {\bibfield
  {journal} {\bibinfo  {journal} {Physical Review Letters}\ }\textbf {\bibinfo
  {volume} {111}},\ \bibinfo {pages} {107204} (\bibinfo {year}
  {2013})}\BibitemShut {NoStop}%
\bibitem [{\citenamefont {Schreier}\ \emph
  {et~al.}(2013{\natexlab{a}})\citenamefont {Schreier}, \citenamefont {Kamra},
  \citenamefont {Weiler}, \citenamefont {Xiao}, \citenamefont {Bauer},
  \citenamefont {Gross},\ and\ \citenamefont
  {Goennenwein}}]{PhysRevB.88.094410}%
  \BibitemOpen
  \bibfield  {author} {\bibinfo {author} {\bibfnamefont {M.}~\bibnamefont
  {Schreier}}, \bibinfo {author} {\bibfnamefont {A.}~\bibnamefont {Kamra}},
  \bibinfo {author} {\bibfnamefont {M.}~\bibnamefont {Weiler}}, \bibinfo
  {author} {\bibfnamefont {J.}~\bibnamefont {Xiao}}, \bibinfo {author}
  {\bibfnamefont {G.~E.~W.}\ \bibnamefont {Bauer}}, \bibinfo {author}
  {\bibfnamefont {R.}~\bibnamefont {Gross}}, \ and\ \bibinfo {author}
  {\bibfnamefont {S.~T.~B.}\ \bibnamefont {Goennenwein}},\ }\href {\doibase
  10.1103/PhysRevB.88.094410} {\bibfield  {journal} {\bibinfo  {journal}
  {Physical Review B}\ }\textbf {\bibinfo {volume} {88}},\ \bibinfo {pages}
  {094410} (\bibinfo {year} {2013}{\natexlab{a}})}\BibitemShut {NoStop}%
\bibitem [{\citenamefont {Bozhko}\ \emph {et~al.}(2016)\citenamefont {Bozhko},
  \citenamefont {Clausen}, \citenamefont {Melkov}, \citenamefont {Victor},
  \citenamefont {Pomyalov}, \citenamefont {Vasyuchka}, \citenamefont {Chumak},
  \citenamefont {Hillebrands},\ and\ \citenamefont {Serga}}]{Bozhko2016}%
  \BibitemOpen
  \bibfield  {author} {\bibinfo {author} {\bibfnamefont {D.~A.}\ \bibnamefont
  {Bozhko}}, \bibinfo {author} {\bibfnamefont {P.}~\bibnamefont {Clausen}},
  \bibinfo {author} {\bibfnamefont {G.~A.}\ \bibnamefont {Melkov}}, \bibinfo
  {author} {\bibfnamefont {S.~L.}\ \bibnamefont {Victor}}, \bibinfo {author}
  {\bibfnamefont {A.}~\bibnamefont {Pomyalov}}, \bibinfo {author}
  {\bibfnamefont {V.~I.}\ \bibnamefont {Vasyuchka}}, \bibinfo {author}
  {\bibfnamefont {A.~V.}\ \bibnamefont {Chumak}}, \bibinfo {author}
  {\bibfnamefont {B.}~\bibnamefont {Hillebrands}}, \ and\ \bibinfo {author}
  {\bibfnamefont {A.~A.}\ \bibnamefont {Serga}},\ }\href@noop {} {\enquote
  {\bibinfo {title} {{Bottleneck accumulation of hybrid magneto-elastic
  bosons}},}\ } (\bibinfo {year} {2016}),\ \Eprint
  {http://arxiv.org/abs/1612.05925} {arXiv:1612.05925} \BibitemShut {NoStop}%
\bibitem [{\citenamefont {Jedema}\ \emph {et~al.}(2001)\citenamefont {Jedema},
  \citenamefont {Filip},\ and\ \citenamefont {van Wees}}]{Jedema2001}%
  \BibitemOpen
  \bibfield  {author} {\bibinfo {author} {\bibfnamefont {F.}~\bibnamefont
  {Jedema}}, \bibinfo {author} {\bibfnamefont {A.}~\bibnamefont {Filip}}, \
  and\ \bibinfo {author} {\bibfnamefont {B.}~\bibnamefont {van Wees}},\ }\href
  {\doibase 10.1038/35066533} {\bibfield  {journal} {\bibinfo  {journal}
  {Nature}\ }\textbf {\bibinfo {volume} {410}},\ \bibinfo {pages} {345}
  (\bibinfo {year} {2001})}\BibitemShut {NoStop}%
\bibitem [{\citenamefont {Lou}\ \emph {et~al.}(2007)\citenamefont {Lou},
  \citenamefont {Adelmann}, \citenamefont {Crooker}, \citenamefont {Garlid},
  \citenamefont {Zhang}, \citenamefont {Reddy}, \citenamefont {Flexner},
  \citenamefont {Palmstrom},\ and\ \citenamefont {Crowell}}]{Lou2006}%
  \BibitemOpen
  \bibfield  {author} {\bibinfo {author} {\bibfnamefont {X.}~\bibnamefont
  {Lou}}, \bibinfo {author} {\bibfnamefont {C.}~\bibnamefont {Adelmann}},
  \bibinfo {author} {\bibfnamefont {S.~A.}\ \bibnamefont {Crooker}}, \bibinfo
  {author} {\bibfnamefont {E.~S.}\ \bibnamefont {Garlid}}, \bibinfo {author}
  {\bibfnamefont {J.}~\bibnamefont {Zhang}}, \bibinfo {author} {\bibfnamefont
  {S.~M.}\ \bibnamefont {Reddy}}, \bibinfo {author} {\bibfnamefont {S.~D.}\
  \bibnamefont {Flexner}}, \bibinfo {author} {\bibfnamefont {C.~J.}\
  \bibnamefont {Palmstrom}}, \ and\ \bibinfo {author} {\bibfnamefont {P.~A.}\
  \bibnamefont {Crowell}},\ }\href {\doibase 10.1038/nphys543} {\bibfield
  {journal} {\bibinfo  {journal} {Nature Physics}\ }\textbf {\bibinfo {volume}
  {3}},\ \bibinfo {pages} {197} (\bibinfo {year} {2007})}\BibitemShut {NoStop}%
\bibitem [{\citenamefont {Tombros}\ \emph {et~al.}(2007)\citenamefont
  {Tombros}, \citenamefont {Jozsa}, \citenamefont {Popinciuc}, \citenamefont
  {Jonkman},\ and\ \citenamefont {van Wees}}]{Tombros2007}%
  \BibitemOpen
  \bibfield  {author} {\bibinfo {author} {\bibfnamefont {N.}~\bibnamefont
  {Tombros}}, \bibinfo {author} {\bibfnamefont {C.}~\bibnamefont {Jozsa}},
  \bibinfo {author} {\bibfnamefont {M.}~\bibnamefont {Popinciuc}}, \bibinfo
  {author} {\bibfnamefont {H.~T.}\ \bibnamefont {Jonkman}}, \ and\ \bibinfo
  {author} {\bibfnamefont {B.~J.}\ \bibnamefont {van Wees}},\ }\href {\doibase
  10.1038/nature06037} {\bibfield  {journal} {\bibinfo  {journal} {Nature}\
  }\textbf {\bibinfo {volume} {448}},\ \bibinfo {pages} {571} (\bibinfo {year}
  {2007})}\BibitemShut {NoStop}%
\bibitem [{\citenamefont {Zutic}\ \emph {et~al.}(2004)\citenamefont {Zutic},
  \citenamefont {Fabian},\ and\ \citenamefont {Sarma}}]{Fabian2004}%
  \BibitemOpen
  \bibfield  {author} {\bibinfo {author} {\bibfnamefont {I.}~\bibnamefont
  {Zutic}}, \bibinfo {author} {\bibfnamefont {J.}~\bibnamefont {Fabian}}, \
  and\ \bibinfo {author} {\bibfnamefont {S.~D.}\ \bibnamefont {Sarma}},\
  }\href@noop {} {\bibfield  {journal} {\bibinfo  {journal} {Reviews of Modern
  Physics}\ }\textbf {\bibinfo {volume} {76}},\ \bibinfo {pages} {323}
  (\bibinfo {year} {2004})}\BibitemShut {NoStop}%
\bibitem [{\citenamefont {Cornelissen}\ \emph {et~al.}(2015)\citenamefont
  {Cornelissen}, \citenamefont {Liu}, \citenamefont {Duine}, \citenamefont
  {Youssef},\ and\ \citenamefont {van Wees}}]{Cornelissen2015}%
  \BibitemOpen
  \bibfield  {author} {\bibinfo {author} {\bibfnamefont {L.~J.}\ \bibnamefont
  {Cornelissen}}, \bibinfo {author} {\bibfnamefont {J.}~\bibnamefont {Liu}},
  \bibinfo {author} {\bibfnamefont {R.~A.}\ \bibnamefont {Duine}}, \bibinfo
  {author} {\bibfnamefont {J.~B.}\ \bibnamefont {Youssef}}, \ and\ \bibinfo
  {author} {\bibfnamefont {B.~J.}\ \bibnamefont {van Wees}},\ }\href {\doibase
  10.1038/nphys3465} {\bibfield  {journal} {\bibinfo  {journal} {Nature
  Physics}\ }\textbf {\bibinfo {volume} {11}},\ \bibinfo {pages} {1022}
  (\bibinfo {year} {2015})}\BibitemShut {NoStop}%
\bibitem [{\citenamefont {Goennenwein}\ \emph {et~al.}(2015)\citenamefont
  {Goennenwein}, \citenamefont {Schlitz}, \citenamefont {Pernpeintner},
  \citenamefont {Ganzhorn}, \citenamefont {Althammer}, \citenamefont {Gross},\
  and\ \citenamefont {Huebl}}]{Goennenwein2015}%
  \BibitemOpen
  \bibfield  {author} {\bibinfo {author} {\bibfnamefont {S.~T.~B.}\
  \bibnamefont {Goennenwein}}, \bibinfo {author} {\bibfnamefont
  {R.}~\bibnamefont {Schlitz}}, \bibinfo {author} {\bibfnamefont
  {M.}~\bibnamefont {Pernpeintner}}, \bibinfo {author} {\bibfnamefont
  {K.}~\bibnamefont {Ganzhorn}}, \bibinfo {author} {\bibfnamefont
  {M.}~\bibnamefont {Althammer}}, \bibinfo {author} {\bibfnamefont
  {R.}~\bibnamefont {Gross}}, \ and\ \bibinfo {author} {\bibfnamefont
  {H.}~\bibnamefont {Huebl}},\ }\href {\doibase 10.1063/1.4935074} {\bibfield
  {journal} {\bibinfo  {journal} {Applied Physics Letters}\ }\textbf {\bibinfo
  {volume} {107}},\ \bibinfo {pages} {172405} (\bibinfo {year}
  {2015})}\BibitemShut {NoStop}%
\bibitem [{\citenamefont {Ganzhorn}\ \emph {et~al.}(2017)\citenamefont
  {Ganzhorn}, \citenamefont {Wimmer}, \citenamefont {Cramer}, \citenamefont
  {Gepr{\"{a}}gs}, \citenamefont {Gross}, \citenamefont {Kl{\"{a}}ui},\ and\
  \citenamefont {Goennenwein}}]{Ganzhorn2017}%
  \BibitemOpen
  \bibfield  {author} {\bibinfo {author} {\bibfnamefont {K.}~\bibnamefont
  {Ganzhorn}}, \bibinfo {author} {\bibfnamefont {T.}~\bibnamefont {Wimmer}},
  \bibinfo {author} {\bibfnamefont {J.}~\bibnamefont {Cramer}}, \bibinfo
  {author} {\bibfnamefont {S.}~\bibnamefont {Gepr{\"{a}}gs}}, \bibinfo {author}
  {\bibfnamefont {R.}~\bibnamefont {Gross}}, \bibinfo {author} {\bibfnamefont
  {M.}~\bibnamefont {Kl{\"{a}}ui}}, \ and\ \bibinfo {author} {\bibfnamefont
  {S.~T.~B.}\ \bibnamefont {Goennenwein}},\ }\href@noop {} {\enquote {\bibinfo
  {title} {{Temperature dependence of the non-local spin Seebeck effect in YIG
  / Pt nanostructures}},}\ } (\bibinfo {year} {2017}),\ \Eprint
  {http://arxiv.org/abs/1701.02635} {arXiv:1701.02635} \BibitemShut {NoStop}%
\bibitem [{\citenamefont {Zhou}\ \emph {et~al.}(2017)\citenamefont {Zhou},
  \citenamefont {Shi}, \citenamefont {Han}, \citenamefont {Yang}, \citenamefont
  {Rao}, \citenamefont {Zhang}, \citenamefont {Zhou}, \citenamefont {Pan},\
  and\ \citenamefont {Song}}]{Zhou2017}%
  \BibitemOpen
  \bibfield  {author} {\bibinfo {author} {\bibfnamefont {X.~J.}\ \bibnamefont
  {Zhou}}, \bibinfo {author} {\bibfnamefont {G.~Y.}\ \bibnamefont {Shi}},
  \bibinfo {author} {\bibfnamefont {J.~H.}\ \bibnamefont {Han}}, \bibinfo
  {author} {\bibfnamefont {Q.~H.}\ \bibnamefont {Yang}}, \bibinfo {author}
  {\bibfnamefont {Y.~H.}\ \bibnamefont {Rao}}, \bibinfo {author} {\bibfnamefont
  {H.~W.}\ \bibnamefont {Zhang}}, \bibinfo {author} {\bibfnamefont {S.~M.}\
  \bibnamefont {Zhou}}, \bibinfo {author} {\bibfnamefont {F.}~\bibnamefont
  {Pan}}, \ and\ \bibinfo {author} {\bibfnamefont {C.}~\bibnamefont {Song}},\
  }\href {\doibase 10.1063/1.4976074} {\bibfield  {journal} {\bibinfo
  {journal} {Applied Physics Letters}\ }\textbf {\bibinfo {volume} {110}},\
  \bibinfo {pages} {062407} (\bibinfo {year} {2017})}\BibitemShut {NoStop}%
\bibitem [{\citenamefont {Cornelissen}\ \emph
  {et~al.}(2016{\natexlab{a}})\citenamefont {Cornelissen}, \citenamefont
  {Shan},\ and\ \citenamefont {van Wees}}]{Cornelissen2016b}%
  \BibitemOpen
  \bibfield  {author} {\bibinfo {author} {\bibfnamefont {L.~J.}\ \bibnamefont
  {Cornelissen}}, \bibinfo {author} {\bibfnamefont {J.}~\bibnamefont {Shan}}, \
  and\ \bibinfo {author} {\bibfnamefont {B.~J.}\ \bibnamefont {van Wees}},\
  }\href {\doibase 10.1103/PhysRevB.94.180402} {\bibfield  {journal} {\bibinfo
  {journal} {Physical Review B}\ }\textbf {\bibinfo {volume} {94}},\ \bibinfo
  {pages} {180402(R)} (\bibinfo {year} {2016}{\natexlab{a}})}\BibitemShut
  {NoStop}%
\bibitem [{\citenamefont {Cornelissen}\ and\ \citenamefont {van
  Wees}(2016)}]{Cornelissen2016}%
  \BibitemOpen
  \bibfield  {author} {\bibinfo {author} {\bibfnamefont {L.~J.}\ \bibnamefont
  {Cornelissen}}\ and\ \bibinfo {author} {\bibfnamefont {B.~J.}\ \bibnamefont
  {van Wees}},\ }\href {\doibase 10.1103/PhysRevB.93.020403} {\bibfield
  {journal} {\bibinfo  {journal} {Physical Review B}\ }\textbf {\bibinfo
  {volume} {93}},\ \bibinfo {pages} {020403(R)} (\bibinfo {year}
  {2016})}\BibitemShut {NoStop}%
\bibitem [{\citenamefont {Chumak}\ \emph {et~al.}(2015)\citenamefont {Chumak},
  \citenamefont {Vasyuchka}, \citenamefont {Serga},\ and\ \citenamefont
  {Hillebrands}}]{Chumak2015}%
  \BibitemOpen
  \bibfield  {author} {\bibinfo {author} {\bibfnamefont {A.~V.}\ \bibnamefont
  {Chumak}}, \bibinfo {author} {\bibfnamefont {V.~I.}\ \bibnamefont
  {Vasyuchka}}, \bibinfo {author} {\bibfnamefont {A.~A.}\ \bibnamefont
  {Serga}}, \ and\ \bibinfo {author} {\bibfnamefont {B.}~\bibnamefont
  {Hillebrands}},\ }\href {\doibase 10.1038/nphys3347} {\bibfield  {journal}
  {\bibinfo  {journal} {Nature Physics}\ }\textbf {\bibinfo {volume} {11}},\
  \bibinfo {pages} {453} (\bibinfo {year} {2015})}\BibitemShut {NoStop}%
\bibitem [{\citenamefont {Ganzhorn}\ \emph {et~al.}(2016)\citenamefont
  {Ganzhorn}, \citenamefont {Klingler}, \citenamefont {Wimmer}, \citenamefont
  {Gepr{\"{a}}gs}, \citenamefont {Gross}, \citenamefont {Huebl},\ and\
  \citenamefont {Goennenwein}}]{Ganzhorn2016a}%
  \BibitemOpen
  \bibfield  {author} {\bibinfo {author} {\bibfnamefont {K.}~\bibnamefont
  {Ganzhorn}}, \bibinfo {author} {\bibfnamefont {S.}~\bibnamefont {Klingler}},
  \bibinfo {author} {\bibfnamefont {T.}~\bibnamefont {Wimmer}}, \bibinfo
  {author} {\bibfnamefont {S.}~\bibnamefont {Gepr{\"{a}}gs}}, \bibinfo {author}
  {\bibfnamefont {R.}~\bibnamefont {Gross}}, \bibinfo {author} {\bibfnamefont
  {H.}~\bibnamefont {Huebl}}, \ and\ \bibinfo {author} {\bibfnamefont
  {S.~T.~B.}\ \bibnamefont {Goennenwein}},\ }\href {\doibase 10.1063/1.4958893}
  {\bibfield  {journal} {\bibinfo  {journal} {Applied Physics Letters}\
  }\textbf {\bibinfo {volume} {109}},\ \bibinfo {pages} {022405} (\bibinfo
  {year} {2016})}\BibitemShut {NoStop}%
\bibitem [{\citenamefont {Uchida}\ \emph
  {et~al.}(2010{\natexlab{a}})\citenamefont {Uchida}, \citenamefont {Xiao},
  \citenamefont {Adachi}, \citenamefont {Ohe}, \citenamefont {Takahashi},
  \citenamefont {Ieda}, \citenamefont {Ota}, \citenamefont {Kajiwara},
  \citenamefont {Umezawa}, \citenamefont {Kawai}, \citenamefont {Bauer},
  \citenamefont {Maekawa},\ and\ \citenamefont {Saitoh}}]{Uchida2010}%
  \BibitemOpen
  \bibfield  {author} {\bibinfo {author} {\bibfnamefont {K.}~\bibnamefont
  {Uchida}}, \bibinfo {author} {\bibfnamefont {J.}~\bibnamefont {Xiao}},
  \bibinfo {author} {\bibfnamefont {H.}~\bibnamefont {Adachi}}, \bibinfo
  {author} {\bibfnamefont {J.}~\bibnamefont {Ohe}}, \bibinfo {author}
  {\bibfnamefont {S.}~\bibnamefont {Takahashi}}, \bibinfo {author}
  {\bibfnamefont {J.}~\bibnamefont {Ieda}}, \bibinfo {author} {\bibfnamefont
  {T.}~\bibnamefont {Ota}}, \bibinfo {author} {\bibfnamefont {Y.}~\bibnamefont
  {Kajiwara}}, \bibinfo {author} {\bibfnamefont {H.}~\bibnamefont {Umezawa}},
  \bibinfo {author} {\bibfnamefont {H.}~\bibnamefont {Kawai}}, \bibinfo
  {author} {\bibfnamefont {G.~E.~W.}\ \bibnamefont {Bauer}}, \bibinfo {author}
  {\bibfnamefont {S.}~\bibnamefont {Maekawa}}, \ and\ \bibinfo {author}
  {\bibfnamefont {E.}~\bibnamefont {Saitoh}},\ }\href {\doibase
  10.1038/nmat2856} {\bibfield  {journal} {\bibinfo  {journal} {Nature
  Materials}\ }\textbf {\bibinfo {volume} {9}},\ \bibinfo {pages} {894}
  (\bibinfo {year} {2010}{\natexlab{a}})}\BibitemShut {NoStop}%
\bibitem [{\citenamefont {Vlietstra}\ \emph {et~al.}(2014)\citenamefont
  {Vlietstra}, \citenamefont {Shan}, \citenamefont {van Wees}, \citenamefont
  {Isasa}, \citenamefont {Casanova},\ and\ \citenamefont {{Ben
  Youssef}}}]{Vlietstra2014}%
  \BibitemOpen
  \bibfield  {author} {\bibinfo {author} {\bibfnamefont {N.}~\bibnamefont
  {Vlietstra}}, \bibinfo {author} {\bibfnamefont {J.}~\bibnamefont {Shan}},
  \bibinfo {author} {\bibfnamefont {B.~J.}\ \bibnamefont {van Wees}}, \bibinfo
  {author} {\bibfnamefont {M.}~\bibnamefont {Isasa}}, \bibinfo {author}
  {\bibfnamefont {F.}~\bibnamefont {Casanova}}, \ and\ \bibinfo {author}
  {\bibfnamefont {J.}~\bibnamefont {{Ben Youssef}}},\ }\href {\doibase
  10.1103/PhysRevB.90.174436} {\bibfield  {journal} {\bibinfo  {journal}
  {Physical Review B}\ }\textbf {\bibinfo {volume} {90}},\ \bibinfo {pages}
  {174436} (\bibinfo {year} {2014})}\BibitemShut {NoStop}%
\bibitem [{\citenamefont {Weiler}\ \emph {et~al.}(2012)\citenamefont {Weiler},
  \citenamefont {Althammer}, \citenamefont {Czeschka}, \citenamefont {Huebl},
  \citenamefont {Wagner}, \citenamefont {Opel}, \citenamefont {Imort},
  \citenamefont {Reiss}, \citenamefont {Thomas}, \citenamefont {Gross},\ and\
  \citenamefont {Goennenwein}}]{Weiler2012}%
  \BibitemOpen
  \bibfield  {author} {\bibinfo {author} {\bibfnamefont {M.}~\bibnamefont
  {Weiler}}, \bibinfo {author} {\bibfnamefont {M.}~\bibnamefont {Althammer}},
  \bibinfo {author} {\bibfnamefont {F.~D.}\ \bibnamefont {Czeschka}}, \bibinfo
  {author} {\bibfnamefont {H.}~\bibnamefont {Huebl}}, \bibinfo {author}
  {\bibfnamefont {M.~S.}\ \bibnamefont {Wagner}}, \bibinfo {author}
  {\bibfnamefont {M.}~\bibnamefont {Opel}}, \bibinfo {author} {\bibfnamefont
  {I.~M.}\ \bibnamefont {Imort}}, \bibinfo {author} {\bibfnamefont
  {G.}~\bibnamefont {Reiss}}, \bibinfo {author} {\bibfnamefont
  {A.}~\bibnamefont {Thomas}}, \bibinfo {author} {\bibfnamefont
  {R.}~\bibnamefont {Gross}}, \ and\ \bibinfo {author} {\bibfnamefont
  {S.~T.~B.}\ \bibnamefont {Goennenwein}},\ }\href {\doibase
  10.1103/PhysRevLett.108.106602} {\bibfield  {journal} {\bibinfo  {journal}
  {Physical Review Letters}\ }\textbf {\bibinfo {volume} {108}},\ \bibinfo
  {pages} {106602} (\bibinfo {year} {2012})}\BibitemShut {NoStop}%
\bibitem [{\citenamefont {Meier}\ \emph {et~al.}(2013)\citenamefont {Meier},
  \citenamefont {Kuschel}, \citenamefont {Shen}, \citenamefont {Gupta},
  \citenamefont {Kikkawa}, \citenamefont {Uchida}, \citenamefont {Saitoh},
  \citenamefont {Schmalhorst},\ and\ \citenamefont {Reiss}}]{Meier2013}%
  \BibitemOpen
  \bibfield  {author} {\bibinfo {author} {\bibfnamefont {D.}~\bibnamefont
  {Meier}}, \bibinfo {author} {\bibfnamefont {T.}~\bibnamefont {Kuschel}},
  \bibinfo {author} {\bibfnamefont {L.}~\bibnamefont {Shen}}, \bibinfo {author}
  {\bibfnamefont {A.}~\bibnamefont {Gupta}}, \bibinfo {author} {\bibfnamefont
  {T.}~\bibnamefont {Kikkawa}}, \bibinfo {author} {\bibfnamefont
  {K.}~\bibnamefont {Uchida}}, \bibinfo {author} {\bibfnamefont
  {E.}~\bibnamefont {Saitoh}}, \bibinfo {author} {\bibfnamefont {J.-M.}\
  \bibnamefont {Schmalhorst}}, \ and\ \bibinfo {author} {\bibfnamefont
  {G.}~\bibnamefont {Reiss}},\ }\href {\doibase 10.1103/PhysRevB.87.054421}
  {\bibfield  {journal} {\bibinfo  {journal} {Physical Review B}\ }\textbf
  {\bibinfo {volume} {87}},\ \bibinfo {pages} {054421} (\bibinfo {year}
  {2013})}\BibitemShut {NoStop}%
\bibitem [{\citenamefont {Uchida}\ \emph
  {et~al.}(2010{\natexlab{b}})\citenamefont {Uchida}, \citenamefont {Adachi},
  \citenamefont {Ota}, \citenamefont {Nakayama}, \citenamefont {Maekawa},\ and\
  \citenamefont {Saitoh}}]{Uchida2010a}%
  \BibitemOpen
  \bibfield  {author} {\bibinfo {author} {\bibfnamefont {K.~I.}\ \bibnamefont
  {Uchida}}, \bibinfo {author} {\bibfnamefont {H.}~\bibnamefont {Adachi}},
  \bibinfo {author} {\bibfnamefont {T.}~\bibnamefont {Ota}}, \bibinfo {author}
  {\bibfnamefont {H.}~\bibnamefont {Nakayama}}, \bibinfo {author}
  {\bibfnamefont {S.}~\bibnamefont {Maekawa}}, \ and\ \bibinfo {author}
  {\bibfnamefont {E.}~\bibnamefont {Saitoh}},\ }\href {\doibase
  10.1063/1.3507386} {\bibfield  {journal} {\bibinfo  {journal} {Applied
  Physics Letters}\ }\textbf {\bibinfo {volume} {97}},\ \bibinfo {pages}
  {172505} (\bibinfo {year} {2010}{\natexlab{b}})}\BibitemShut {NoStop}%
\bibitem [{\citenamefont {Schreier}\ \emph
  {et~al.}(2013{\natexlab{b}})\citenamefont {Schreier}, \citenamefont
  {Roschewsky}, \citenamefont {Dobler}, \citenamefont {Meyer}, \citenamefont
  {Huebl}, \citenamefont {Gross},\ and\ \citenamefont
  {Goennenwein}}]{Schreier2013}%
  \BibitemOpen
  \bibfield  {author} {\bibinfo {author} {\bibfnamefont {M.}~\bibnamefont
  {Schreier}}, \bibinfo {author} {\bibfnamefont {N.}~\bibnamefont
  {Roschewsky}}, \bibinfo {author} {\bibfnamefont {E.}~\bibnamefont {Dobler}},
  \bibinfo {author} {\bibfnamefont {S.}~\bibnamefont {Meyer}}, \bibinfo
  {author} {\bibfnamefont {H.}~\bibnamefont {Huebl}}, \bibinfo {author}
  {\bibfnamefont {R.}~\bibnamefont {Gross}}, \ and\ \bibinfo {author}
  {\bibfnamefont {S.~T.~B.}\ \bibnamefont {Goennenwein}},\ }\href {\doibase
  10.1063/1.4839395} {\bibfield  {journal} {\bibinfo  {journal} {Applied
  Physics Letters}\ }\textbf {\bibinfo {volume} {103}},\ \bibinfo {pages}
  {242404} (\bibinfo {year} {2013}{\natexlab{b}})}\BibitemShut {NoStop}%
\bibitem [{\citenamefont {Shan}\ \emph {et~al.}(2016)\citenamefont {Shan},
  \citenamefont {Cornelissen}, \citenamefont {Vlietstra}, \citenamefont {{Ben
  Youssef}}, \citenamefont {Kuschel}, \citenamefont {Duine},\ and\
  \citenamefont {van Wees}}]{Shan}%
  \BibitemOpen
  \bibfield  {author} {\bibinfo {author} {\bibfnamefont {J.}~\bibnamefont
  {Shan}}, \bibinfo {author} {\bibfnamefont {L.~J.}\ \bibnamefont
  {Cornelissen}}, \bibinfo {author} {\bibfnamefont {N.}~\bibnamefont
  {Vlietstra}}, \bibinfo {author} {\bibfnamefont {J.}~\bibnamefont {{Ben
  Youssef}}}, \bibinfo {author} {\bibfnamefont {T.}~\bibnamefont {Kuschel}},
  \bibinfo {author} {\bibfnamefont {R.~A.}\ \bibnamefont {Duine}}, \ and\
  \bibinfo {author} {\bibfnamefont {B.~J.}\ \bibnamefont {van Wees}},\
  }\href@noop {} {\bibfield  {journal} {\bibinfo  {journal} {Physical Review
  B}\ }\textbf {\bibinfo {volume} {94}},\ \bibinfo {pages} {174437} (\bibinfo
  {year} {2016})}\BibitemShut {NoStop}%
\bibitem [{\citenamefont {Cornelissen}\ \emph
  {et~al.}(2016{\natexlab{b}})\citenamefont {Cornelissen}, \citenamefont
  {Peters}, \citenamefont {Duine}, \citenamefont {Bauer},\ and\ \citenamefont
  {van Wees}}]{Cornelissen2016a}%
  \BibitemOpen
  \bibfield  {author} {\bibinfo {author} {\bibfnamefont {L.~J.}\ \bibnamefont
  {Cornelissen}}, \bibinfo {author} {\bibfnamefont {K.~J.~H.}\ \bibnamefont
  {Peters}}, \bibinfo {author} {\bibfnamefont {R.~A.}\ \bibnamefont {Duine}},
  \bibinfo {author} {\bibfnamefont {G.~E.~W.}\ \bibnamefont {Bauer}}, \ and\
  \bibinfo {author} {\bibfnamefont {B.~J.}\ \bibnamefont {van Wees}},\ }\href
  {\doibase 10.1103/PhysRevB.94.014412} {\bibfield  {journal} {\bibinfo
  {journal} {Physical Review B}\ }\textbf {\bibinfo {volume} {94}},\ \bibinfo
  {pages} {014412} (\bibinfo {year} {2016}{\natexlab{b}})}\BibitemShut
  {NoStop}%
\bibitem [{\citenamefont {Saitoh}\ \emph {et~al.}(2006)\citenamefont {Saitoh},
  \citenamefont {Ueda}, \citenamefont {Miyajima},\ and\ \citenamefont
  {Tatara}}]{Saitoh2006}%
  \BibitemOpen
  \bibfield  {author} {\bibinfo {author} {\bibfnamefont {E.}~\bibnamefont
  {Saitoh}}, \bibinfo {author} {\bibfnamefont {M.}~\bibnamefont {Ueda}},
  \bibinfo {author} {\bibfnamefont {H.}~\bibnamefont {Miyajima}}, \ and\
  \bibinfo {author} {\bibfnamefont {G.}~\bibnamefont {Tatara}},\ }\href
  {\doibase 10.1063/1.2199473} {\bibfield  {journal} {\bibinfo  {journal}
  {Applied Physics Letters}\ }\textbf {\bibinfo {volume} {88}},\ \bibinfo
  {pages} {182509} (\bibinfo {year} {2006})}\BibitemShut {NoStop}%
\bibitem [{\citenamefont {Lecraw}\ and\ \citenamefont
  {Walker}(1961)}]{Lecraw1961}%
  \BibitemOpen
  \bibfield  {author} {\bibinfo {author} {\bibfnamefont {R.~C.}\ \bibnamefont
  {Lecraw}}\ and\ \bibinfo {author} {\bibfnamefont {L.~R.}\ \bibnamefont
  {Walker}},\ }\href {\doibase 10.1063/1.2000390} {\bibfield  {journal}
  {\bibinfo  {journal} {Journal of Applied Physics}\ }\textbf {\bibinfo
  {volume} {32}},\ \bibinfo {pages} {S167} (\bibinfo {year}
  {1961})}\BibitemShut {NoStop}%
\bibitem [{\citenamefont {Solt}(1962)}]{Solt1962}%
  \BibitemOpen
  \bibfield  {author} {\bibinfo {author} {\bibfnamefont {I.~H.}\ \bibnamefont
  {Solt}},\ }\href {\doibase 10.1063/1.1728651} {\bibfield  {journal} {\bibinfo
   {journal} {Journal of Applied Physics}\ }\textbf {\bibinfo {volume} {33}},\
  \bibinfo {pages} {1189} (\bibinfo {year} {1962})}\BibitemShut {NoStop}%
\bibitem [{\citenamefont {Daudin}\ \emph {et~al.}(1982)\citenamefont {Daudin},
  \citenamefont {Lagnier},\ and\ \citenamefont {Salce}}]{Daudin1982}%
  \BibitemOpen
  \bibfield  {author} {\bibinfo {author} {\bibfnamefont {B.}~\bibnamefont
  {Daudin}}, \bibinfo {author} {\bibfnamefont {R.}~\bibnamefont {Lagnier}}, \
  and\ \bibinfo {author} {\bibfnamefont {B.}~\bibnamefont {Salce}},\ }\href
  {\doibase 10.1016/0304-8853(82)90092-0} {\bibfield  {journal} {\bibinfo
  {journal} {Journal of Magnetism and Magnetic Materials}\ }\textbf {\bibinfo
  {volume} {27}},\ \bibinfo {pages} {315} (\bibinfo {year} {1982})}\BibitemShut
  {NoStop}%
\bibitem [{\citenamefont {Boona}\ and\ \citenamefont
  {Heremans}(2014)}]{PhysRevB.90.064421}%
  \BibitemOpen
  \bibfield  {author} {\bibinfo {author} {\bibfnamefont {S.~R.}\ \bibnamefont
  {Boona}}\ and\ \bibinfo {author} {\bibfnamefont {J.~P.}\ \bibnamefont
  {Heremans}},\ }\href {\doibase 10.1103/PhysRevB.90.064421} {\bibfield
  {journal} {\bibinfo  {journal} {Physical Review B}\ }\textbf {\bibinfo
  {volume} {90}},\ \bibinfo {pages} {064421} (\bibinfo {year}
  {2014})}\BibitemShut {NoStop}%
\bibitem [{\citenamefont {Xiao}\ \emph {et~al.}(2010)\citenamefont {Xiao},
  \citenamefont {Bauer}, \citenamefont {Uchida}, \citenamefont {Saitoh},\ and\
  \citenamefont {Maekawa}}]{Xiao2010}%
  \BibitemOpen
  \bibfield  {author} {\bibinfo {author} {\bibfnamefont {J.}~\bibnamefont
  {Xiao}}, \bibinfo {author} {\bibfnamefont {G.~E.~W.}\ \bibnamefont {Bauer}},
  \bibinfo {author} {\bibfnamefont {K.-i.}\ \bibnamefont {Uchida}}, \bibinfo
  {author} {\bibfnamefont {E.}~\bibnamefont {Saitoh}}, \ and\ \bibinfo {author}
  {\bibfnamefont {S.}~\bibnamefont {Maekawa}},\ }\href {\doibase
  10.1103/PhysRevB.81.214418} {\bibfield  {journal} {\bibinfo  {journal}
  {Physical Review B}\ }\textbf {\bibinfo {volume} {81}},\ \bibinfo {pages}
  {214418} (\bibinfo {year} {2010})}\BibitemShut {NoStop}%
\bibitem [{\citenamefont {Adachi}\ \emph {et~al.}(2013)\citenamefont {Adachi},
  \citenamefont {Uchida}, \citenamefont {Saitoh},\ and\ \citenamefont
  {Maekawa}}]{Adachi2013}%
  \BibitemOpen
  \bibfield  {author} {\bibinfo {author} {\bibfnamefont {H.}~\bibnamefont
  {Adachi}}, \bibinfo {author} {\bibfnamefont {K.-i.}\ \bibnamefont {Uchida}},
  \bibinfo {author} {\bibfnamefont {E.}~\bibnamefont {Saitoh}}, \ and\ \bibinfo
  {author} {\bibfnamefont {S.}~\bibnamefont {Maekawa}},\ }\href
  {http://stacks.iop.org/0034-4885/76/i=3/a=036501} {\bibfield  {journal}
  {\bibinfo  {journal} {Reports on Progress in Physics}\ }\textbf {\bibinfo
  {volume} {76}},\ \bibinfo {pages} {36501} (\bibinfo {year}
  {2013})}\BibitemShut {NoStop}%
\bibitem [{\citenamefont {Rezende}\ \emph {et~al.}(2014)\citenamefont
  {Rezende}, \citenamefont {Rodr{\'{i}}guez-Su{\'{a}}rez}, \citenamefont
  {Cunha}, \citenamefont {Rodrigues}, \citenamefont {Machado}, \citenamefont
  {{Fonseca Guerra}}, \citenamefont {{Lopez Ortiz}},\ and\ \citenamefont
  {Azevedo}}]{PhysRevB.89.014416}%
  \BibitemOpen
  \bibfield  {author} {\bibinfo {author} {\bibfnamefont {S.~M.}\ \bibnamefont
  {Rezende}}, \bibinfo {author} {\bibfnamefont {R.~L.}\ \bibnamefont
  {Rodr{\'{i}}guez-Su{\'{a}}rez}}, \bibinfo {author} {\bibfnamefont {R.~O.}\
  \bibnamefont {Cunha}}, \bibinfo {author} {\bibfnamefont {A.~R.}\ \bibnamefont
  {Rodrigues}}, \bibinfo {author} {\bibfnamefont {F.~L.~A.}\ \bibnamefont
  {Machado}}, \bibinfo {author} {\bibfnamefont {G.~A.}\ \bibnamefont {{Fonseca
  Guerra}}}, \bibinfo {author} {\bibfnamefont {J.~C.}\ \bibnamefont {{Lopez
  Ortiz}}}, \ and\ \bibinfo {author} {\bibfnamefont {A.}~\bibnamefont
  {Azevedo}},\ }\href {\doibase 10.1103/PhysRevB.89.014416} {\bibfield
  {journal} {\bibinfo  {journal} {Physical Review B}\ }\textbf {\bibinfo
  {volume} {89}},\ \bibinfo {pages} {014416} (\bibinfo {year}
  {2014})}\BibitemShut {NoStop}%
\bibitem [{\citenamefont {Slack}\ and\ \citenamefont
  {Oliver}(1971)}]{Slack1971}%
  \BibitemOpen
  \bibfield  {author} {\bibinfo {author} {\bibfnamefont {G.~A.}\ \bibnamefont
  {Slack}}\ and\ \bibinfo {author} {\bibfnamefont {D.~W.}\ \bibnamefont
  {Oliver}},\ }\href {\doibase 10.1103/PhysRevB.4.592} {\bibfield  {journal}
  {\bibinfo  {journal} {Physical Review B}\ }\textbf {\bibinfo {volume} {4}},\
  \bibinfo {pages} {592} (\bibinfo {year} {1971})}\BibitemShut {NoStop}%
\bibitem [{\citenamefont {Kikkawa}\ \emph {et~al.}(2015)\citenamefont
  {Kikkawa}, \citenamefont {Uchida}, \citenamefont {Daimon}, \citenamefont
  {Qiu}, \citenamefont {Shiomi},\ and\ \citenamefont {Saitoh}}]{Kikkawa2015}%
  \BibitemOpen
  \bibfield  {author} {\bibinfo {author} {\bibfnamefont {T.}~\bibnamefont
  {Kikkawa}}, \bibinfo {author} {\bibfnamefont {K.-i.}\ \bibnamefont {Uchida}},
  \bibinfo {author} {\bibfnamefont {S.}~\bibnamefont {Daimon}}, \bibinfo
  {author} {\bibfnamefont {Z.}~\bibnamefont {Qiu}}, \bibinfo {author}
  {\bibfnamefont {Y.}~\bibnamefont {Shiomi}}, \ and\ \bibinfo {author}
  {\bibfnamefont {E.}~\bibnamefont {Saitoh}},\ }\href {\doibase
  10.1103/PhysRevB.92.064413} {\bibfield  {journal} {\bibinfo  {journal}
  {Physical Review B}\ }\textbf {\bibinfo {volume} {92}},\ \bibinfo {pages}
  {064413} (\bibinfo {year} {2015})}\BibitemShut {NoStop}%
\bibitem [{\citenamefont {Jin}\ \emph {et~al.}(2015)\citenamefont {Jin},
  \citenamefont {Boona}, \citenamefont {Yang}, \citenamefont {Myers},\ and\
  \citenamefont {Heremans}}]{Jin2015}%
  \BibitemOpen
  \bibfield  {author} {\bibinfo {author} {\bibfnamefont {H.}~\bibnamefont
  {Jin}}, \bibinfo {author} {\bibfnamefont {S.~R.}\ \bibnamefont {Boona}},
  \bibinfo {author} {\bibfnamefont {Z.}~\bibnamefont {Yang}}, \bibinfo {author}
  {\bibfnamefont {R.~C.}\ \bibnamefont {Myers}}, \ and\ \bibinfo {author}
  {\bibfnamefont {J.~P.}\ \bibnamefont {Heremans}},\ }\href
  {http://arxiv.org/pdf/1504.00895.pdf} {\bibfield  {journal} {\bibinfo
  {journal} {Physical Review B}\ }\textbf {\bibinfo {volume} {92}},\ \bibinfo
  {pages} {054436} (\bibinfo {year} {2015})}\BibitemShut {NoStop}%
\bibitem [{\citenamefont {Kikkawa}\ \emph
  {et~al.}(2016{\natexlab{b}})\citenamefont {Kikkawa}, \citenamefont {Uchida},
  \citenamefont {Daimon},\ and\ \citenamefont {Saitoh}}]{Kikkawa2016a}%
  \BibitemOpen
  \bibfield  {author} {\bibinfo {author} {\bibfnamefont {T.}~\bibnamefont
  {Kikkawa}}, \bibinfo {author} {\bibfnamefont {K.~I.}\ \bibnamefont {Uchida}},
  \bibinfo {author} {\bibfnamefont {S.}~\bibnamefont {Daimon}}, \ and\ \bibinfo
  {author} {\bibfnamefont {E.}~\bibnamefont {Saitoh}},\ }\href {\doibase
  10.7566/JPSJ.85.065003} {\bibfield  {journal} {\bibinfo  {journal} {Journal
  of the Physical Society of Japan}\ }\textbf {\bibinfo {volume} {85}},\
  \bibinfo {pages} {065003} (\bibinfo {year} {2016}{\natexlab{b}})}\BibitemShut
  {NoStop}%
\bibitem [{\citenamefont {Bakker}\ \emph {et~al.}(2010)\citenamefont {Bakker},
  \citenamefont {Slachter}, \citenamefont {Adam},\ and\ \citenamefont {van
  Wees}}]{Bakker2010}%
  \BibitemOpen
  \bibfield  {author} {\bibinfo {author} {\bibfnamefont {F.~L.}\ \bibnamefont
  {Bakker}}, \bibinfo {author} {\bibfnamefont {A.}~\bibnamefont {Slachter}},
  \bibinfo {author} {\bibfnamefont {J.-P.}\ \bibnamefont {Adam}}, \ and\
  \bibinfo {author} {\bibfnamefont {B.~J.}\ \bibnamefont {van Wees}},\ }\href
  {\doibase 10.1103/PhysRevLett.105.136601} {\bibfield  {journal} {\bibinfo
  {journal} {Physical Review Letters}\ }\textbf {\bibinfo {volume} {105}},\
  \bibinfo {pages} {136601} (\bibinfo {year} {2010})}\BibitemShut {NoStop}%
\end{thebibliography}%

\end{document}